\documentclass[%
aip,
jmp,
amsmath,amssymb,
reprint,%
author-year,%
author-numerical,%
]{revtex4-2}
\usepackage{graphicx}
\usepackage{dcolumn}
\usepackage{bm}
\usepackage{booktabs}
\usepackage{multirow}
\usepackage{bigints}
\usepackage{enumerate} 
\usepackage{units}
\usepackage{tabularx}
\usepackage{makecell}
 
\usepackage{rotating} 
\usepackage[paper=portrait,pagesize]{typearea}
\numberwithin{equation}{section}
\DeclareUnicodeCharacter{0301}{U+0027} 
\DeclareUnicodeCharacter{2212}{-}
\begin{document}
 
\title{The Earth’s long-term climate changes and ice ages:
a derivation of ``Milankovitch cycles" from first principles}
 
\author{R. C. T. Rainey} 
\affiliation{Rod Rainey \& Associates Ltd., Bath, UK}
 \altaffiliation[Also at ]{School of Engineering, University of Southampton}  \email{enquiries@RRandA.co.uk, r.rainey@soton.ac.uk}

\date{\today} 
\begin{abstract}
Long-term changes in the tilt of the Earth’s axis, relative to the plane of its orbit, are of great significance to long-term climate change, because they control the size of the arctic and Antarctic circles. These ``Milankovitch cycles" have generally been calculated by numerical integration of Newton’s equations of motion, and there is some controversy over the results because they are sensitive to numerical drift over the very long computer simulations involved. In this paper the cycles are calculated from first principles, without any reliance on computer simulation. The problem is one of planetary precession, and is solvable by the methods used to study the precession of a spinning top. It is shown that the main component of ``Milankovitch cycles" has a period of 41,000 years and is due to one of the modes of precession of the Earth-Venus system. The other mode of this system produces a component of period 29,500 years, and a third component of period 54,000 years results from the influence of the precession of the orbits of Jupiter and Saturn.

These results agree closely with several of the numerical simulations in the literature, and strongly suggest that other different results in the literature are incorrect. 

\end{abstract}

\maketitle

\section{Introduction}\label{intro}
\begin{figure}
	\centering\includegraphics[scale = 0.8]{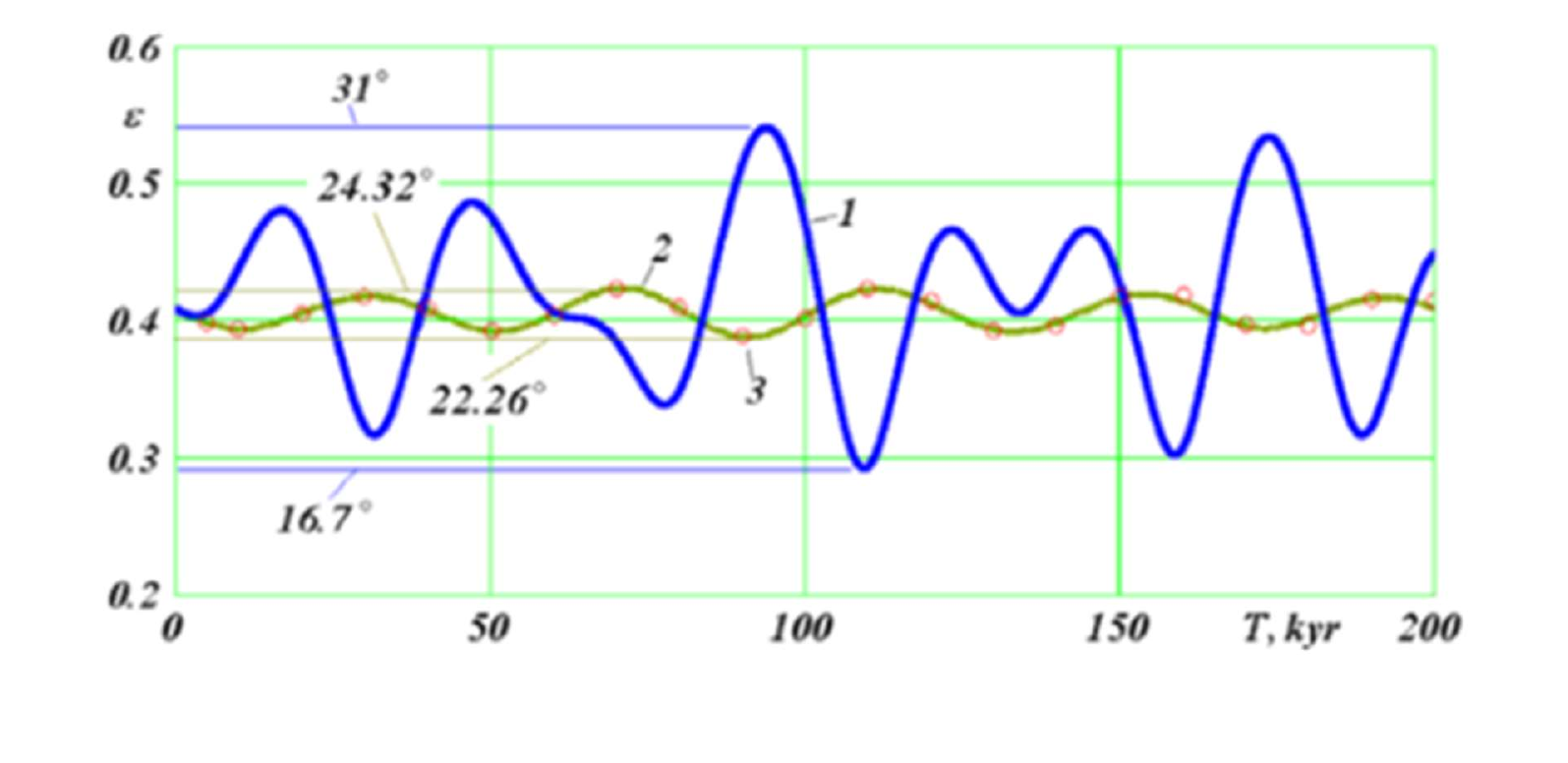}
	\caption{(taken from \citet[Fig.11]{smulsky2016fundamental}) The predicted tilt of the Earth's spin axis, relative to the plane of its orbit (i.e. its obliquity $\varepsilon$, in radians) over the next 200,000 years. All results are from numerical integration: 1, 2, and 3 are from \citet{smulsky2016fundamental}, \citet{laskar2004long} and \citet{ Sharaf:1969aa} respectively.}\label{fig1}
\end{figure}

Long-term variations in the tilt of the Earth’s spin axis, relative to the plane of its orbit, are of minor astronomical significance, but they are of great importance to long-term climate change. They result in variations in the size of the arctic and Antarctic circles, and are a widely-accepted explanation (e.g. \citet{imbrie1986ice, muller2002ice}) for the regular sequence of ice ages over the past 2 million years, revealed by seabed cores.
There is some controversy in the literature over how large these tilt variations are, and their periods of variation.  Figure \ref{fig1} is taken from a recent paper \citet{ smulsky2016fundamental} and shows computed tilt variations many times larger than previous computations \citet{ laskar2004long, Sharaf:1969aa} also shown in Figure \ref{fig1}.

The results in Figure \ref{fig1} are all from numerical integration; the large differences between them appear to be caused by numerical drift over the very long integration times. An important such numerical integration is by  \textcite{ quinn1991three}, which makes use of a numerical model developed by NASA for the exacting requirements of space travel. The variations in the orbital planes of all the planets, over 3 million years, are considered relative to a fixed plane. The results are presented (\citet[Figure 6]{ quinn1991three}) as polar plots of tilt magnitude against tilt orientation (ascending node). Particularly striking are the plots for Jupiter and Saturn, which are essentially circular, suggesting that the mechanism is the same as that which causes the circular precession of the axis of a spinning top. This mechanism is described on p.213 of the well-known textbook \cite{kibble2004classical} by Kibble and Berkshire; in this paper we will rely extensively on that textbook for notation, methods of analysis, and detailed explanations.

The plane of the orbit of the Earth is also seen in Figure 6 of \textcite{ quinn1991three} to be precessing, but in a more complicated way, perhaps related to the similar precession of the orbit of Venus. The Earth’s orbital precession is clearly relevant to changes in the tilt of the Earth’s spin axis, relative to its orbit. In this paper we will seek a physical explanation, and derive the long-term variations of the tilt of the Earth’s spin axis relative to its orbit without recourse to numerical integration, using the methods described in Kibble and Berkshire’s textbook. Our analysis will of necessity be approximate and will only be valid for the relatively recent past (in the very long term, the behavior of the solar system is extremely complicated, see \textcite[section 14.4]{kibble2004classical} ) but it will be easily accurate enough to resolve the controversy illustrated in \mbox{Figure \ref{fig1}}.

The notation in this paper follows a systematic scheme, shown in Table \ref{tablea1.1a} and Table \ref{tablea1.1b} in \mbox{Appendix \ref{app1}}.
\section{Precession of the orbit of the Moon}\label{moon}
It is convenient to begin by considering the simple case of the precession of the plane of the orbit of the Moon around the Earth, because there the precession period is well-known, and gives a useful check on the accuracy of our calculation method, which will be used repeatedly. The case is also important in its own right, because the Moon is the most important influence on the precession of the Earth’s spin axis.

The precession of the orbit of the Moon is explained qualitatively by  \citet{kibble2004classical}(p.144) as being due to the gradient in the Sun’s gravitational field. They define (p.130) a gravitational field as being the vector acceleration $\mathbf{g(x)}$ imparted by the field to a particle at position $\mathbf{x}$, but do not discuss the properties of its gradient. If $\mathbf{g}$ has components $g_x , g_y, g_z $ in orthogonal directions $x, y, z$ then a small change $ \delta  \mathbf{x}$ in position $\mathbf{x}$ with components $\delta x _x , \delta x _y ,  \delta x_z $ will produce a small change $ \delta  \mathbf{g}$ in $\mathbf{g}$ with components:
\begin{equation}\label{eq:2.1}
	\begin{bmatrix}
		\delta  g_x \\
		\\
		\delta  g_y \\
		\\
		\delta  g_z
	\end{bmatrix}
	=
	\begin{bmatrix}
		 \frac{\partial g_x}{\partial x} \frac{\partial g_x}{\partial y} \frac{\partial g_x}{\partial z} \\\\
		 \frac{\partial g_y}{\partial x} \frac{\partial g_y}{\partial y} \frac{\partial g_y}{\partial z} \\\\
		 \frac{\partial g_z}{\partial x} \frac{\partial g_z}{\partial y} \frac{\partial g_z}{\partial z} 
	\end{bmatrix}
	\begin{bmatrix}
		\delta  x_x \\\\
		\delta  x_y \\\\
		\delta  x_z
	\end{bmatrix}
\end{equation} 
Following  the tensor notation of \textcite{kibble2004classical} (p.401) we will write this as:
\begin{equation}\label{eq:2.2}
	\delta  \mathbf{g} = \mathbf{ \mathsf{C} \delta  \bm{x}} 
\end{equation}
where the gradient of the Sun’s gravitational field is written as the tensor $ \mathsf{C}$. In their general discussion of tensors, \textcite{kibble2004classical} (p.401) distinguish the special case of a symmetric tensor, and $ \mathsf{C}$ is such a case because \citep[eqn. 6.3]{kibble2004classical} $\mathbf{g}$ is the gradient of a potential $\Phi$ so that:
\begin{equation*}\label{eq:2.3}
	\frac{\partial g_x}{\partial y}= \frac{\partial}{\partial y} \frac{\partial \Phi}{\partial x} = \frac{\partial}{\partial x} \frac{\partial \Phi}{\partial y} = \frac{\partial g_y}{\partial x}
	\quad \text{etc.}
\end{equation*}
 
We conclude (\citet[ p.404]{kibble2004classical}) that the orthogonal directions $x, y, z$ can be chosen so that the matrix in  (\ref{eq:2.1}) diagonalizes. Moreover, if these ``principal directions" are chosen, then the sum of the diagonal elements of the matrix in (\ref{eq:2.1})  is:
\begin{equation}\label{eq:2.4}
	\frac{\partial^2 \Phi}{\partial x^2} +\frac{\partial^2 \Phi}{\partial y^2} +\frac{\partial^2 \Phi}{\partial z^2} = 0
\end{equation} 
by definition of a potential (Laplace’s equation \cite[eqn. 6.49]{kibble2004classical}). We now make use of these properties of $ \mathsf{C}$.

We can take a frame of reference which is centered on the combined center of gravity $C$ of the Earth and Moon, but does not rotate relative to distant stars. In this frame both the Moon and Sun appear to be in orbit; for the moment we will assume the plane of the Sun’s orbit to be fixed (i.e. we assume the normal $\mathbf{S}$ to the plane of the Sun’s orbit is aligned with the normal $\mathbf{L}$ to the Laplace invariable plane, which is \citet{souami2012solar} the plane normal to the (fixed) total angular momentum of the solar system). For present purposes we can take both orbits to be circular (in fact their mean eccentricities are 0.055 and 0.0167 respectively\cite{Williams:ac, Williams:2020aa}). The radius of the Moon’s orbit is only 0.00257 times that of the Sun, so for positions $\mathbf{x}$ relative to $C$ that are within the Moon’s orbit, we can take the Sun’s gravitational field as:
\begin{equation}\label{eq:2.5}
\mathbf{g}(C) + \mathbf{ \mathsf{C} x}
\end{equation} 
where $\mathbf{g}(C)$ is its value at $C$, and $  \mathsf{C}$ is its gradient there as just defined. If $\mathbf{r}$ and $\mathbf{r^*}$ are the positions of the CGs of the Moon and Earth relative to $C$, and $m$ and $m^*$ are their masses, then the total gravitational force on them from the Sun is:
\begin{equation}\label{eq:2.6}
	m\left(
		\mathbf{g}(C) +  \mathsf{C}\mathbf{r}
	\right) + 
	m^*\left(
		\mathbf{g}(C) +  \mathsf{C}\mathbf{r^*}
		\right)
		 =
		  (m + m^*)\mathbf{g}(C) +  \mathsf{C}(m\mathbf{r} + m^*\mathbf{r^*})  
\end{equation}
	
But $m\mathbf{r} + m^*\mathbf{r^*} = 0$ since $C$ is the center of gravity of $m$ and $m^*$ combined, so the second term on the RHS of  (\ref{eq:2.6}) vanishes. The total gravitational force on $m$ and $m^*$ combined is thus the same as it would be if $ \mathsf{C}$ were zero. Our frame is thus accelerating towards the Sun in the same way that it would if the Sun’s gravitational field was everywhere equal to its value at $C$.

In these latter circumstances of a uniform gravitational field, its effects are indistinguishable from an acceleration of the frame of reference (\citet[p.192]{kibble2004classical}). Thus in our case the effect of the uniform part of the Sun’s gravitational field is cancelled out by the effect of the acceleration of our frame. We can therefore proceed as if our frame centered on $C$ were fixed, provided we ignore the uniform part of the Sun’s gravitational field, equal to its value at $C$. Specifically, we can consider the Moon on its own, and equate the rate-of-change of its angular momentum about $C$, to the moment about $C$ of the forces acting on it, excluding the forces from the first term in (\ref{eq:2.5}).

The gravitational force on the Moon from the Earth acts though $C$ and so has zero moment. However, the second term in (\ref{eq:2.5}), i.e. the effect of gradient $ \mathsf{C}$ of the Sun’s gravitational field, produces a moment about $C$. It is this which causes the precession of the Moon’s orbit, as explained qualitatively by Kibble and Berkshire.

The period of the precession is much longer than the orbital period of the Earth and the Moon, so we are concerned with the long-term average value of the moment. Rather than calculate the moment at each position of the Sun and Moon, and take the average (as suggested in \citet[ p.157]{kibble2004classical}), we can adopt the procedure introduced by Newton (as described in \citet[p.531]{cohen1999principia}) of smearing the mass of the Sun over its orbit seen in our frame, and likewise the Moon. We are therefore concerned with the gravitational moment about $C$ produced by one ring of mass $M$ and radius $R$ corresponding to the Sun, on a much smaller ring at its center, of mass $m$ and radius $r$, corresponding to the Moon. Both rings are centered on $C$. 
\begin{sidewaysfigure}
 
	\includegraphics[scale = 0.9]{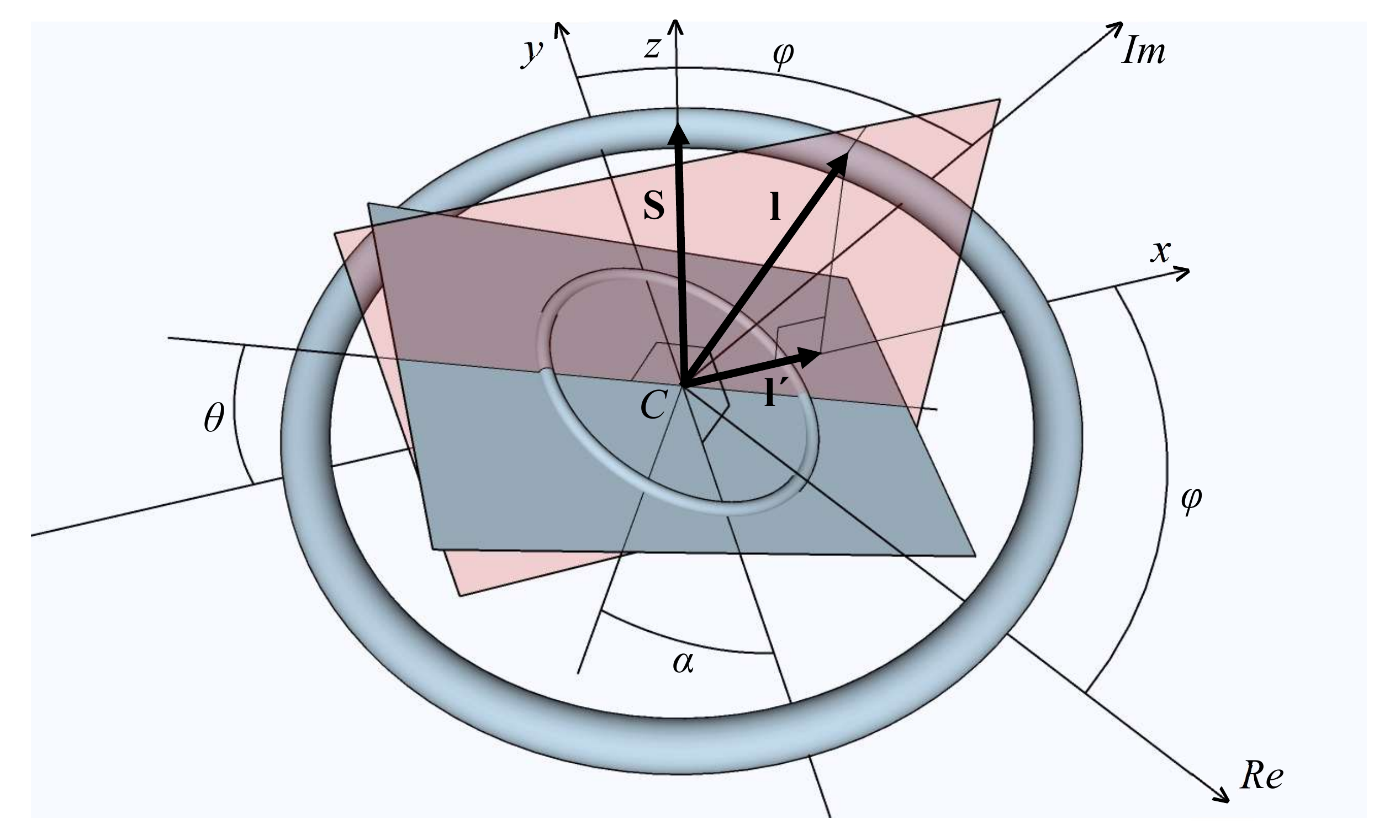}
	\caption{Sun and Moon rings (view from above the Earth's northern hemisphere)}\label{fig2} 
\end{sidewaysfigure}

If both rings were in the same plane the moment would be zero, by symmetry. However, they are not: there is an angle $\theta$ between them of approximately 5\text{\textdegree}, see \mbox{Figure \ref{fig2}} (where it, and the relative size of the rings, is exaggerated for clarity). This produces a moment arising from the gradient $ \mathsf{C}$ in the gravitational force from the Sun’s ring. The principal axes of $ \mathsf{C}$ are, by symmetry, normal to the plane of the Sun’s orbit (direction $z$ in \mbox{Figure \ref{fig2}}), and in any two orthogonal directions in it, which we can choose as the direction of the common diameter of the two rings, and perpendicular to it. These are respectively directions $y$ and $x$ in \mbox{Figure \ref{fig2}}, and are not fixed because the plane of orbit of the moon is precessing. A fixed direction in the $x$-$y$ plane (later denoted as the real axis) is also shown in \mbox{Figure \ref{fig2}}: the $x$-axis is at an angle $\varphi$ to it, which changes with time. The principal component of the gradient in the $z$ direction is easily calculated. If we move a distance $\delta $ from $C$ in the $z$-direction, the gravitational forces from each part of the Sun’s ring are no
 longer in its plane, but an angle $\delta /R$ to it. Since the gravitational force from the Sun is $GM/R^2$ per unit mass, where $G$ is the gravitational constant, we conclude that the principal component of $ \mathsf{C}$ in the $z$-direction is:
\begin{equation}\label{eq:2.7}
-\frac{GM}{R^3} . 
\end{equation} 
The other two principal components of $ \mathsf{C}$ are equal by symmetry, and thus must be equal to:
\begin{equation}\label{eq:2.8}
\frac{GM}{2R^3}
\end{equation}
 because the sum of all three principal components is zero (\ref{eq:2.4}).
 
We now wish to find the moment on the Moon ring produced by $ \mathsf{C}$ . By symmetry, it can have no components about the $z$ and $x$ axes; the moment thus acts entirely about the $y$-axis. The contributions to the moment from each of the principal components of $ \mathsf{C}$  need to be considered in turn. The simplest of the three to consider is the principal component of $ \mathsf{C}$  in the $z$-direction (\ref{eq:2.7}). We first consider the effect of this principal component on a single point on the Moon’s ring, of point mass $\delta  m$. If we denote angular position on the ring, measured round from the common diameter of the two rings, as $\alpha$, see \mbox{Figure \ref{fig2}}, then the $z$-coordinate of this point is:
\begin{equation*}\label{eq:2.9}
r \sin\alpha \sin\theta 
\end{equation*}
The principal component (\ref{eq:2.7}) thus produces a force on $\delta  m$ of:
\begin{equation*}\label{eq:2.10}
 \delta  m \frac{GM}{R^3}r \sin\alpha \sin\theta 
\end{equation*}
which acts in the sense of negative $z$ since the sign of (\ref{eq:2.7}) is negative. Its lever arm about the $y$-axis is
\begin{equation*}\label{eq:2.11}
r\sin\alpha\,\cos\theta = r\sin\alpha 
\end{equation*}
 since $\cos\theta \approx1$ for our small $\theta$. Thus it will produce a moment about the $y$-axis, in the sense to reduce $\theta$, of:
\begin{equation*}\label{eq:2.12}
	\delta m \frac{GM}{R^3}r^2 \sin^2\alpha\sin\theta.  
\end{equation*} 
The total moment on the Moon’s ring will be the sum of the contributions from all the masses $\delta  m$ around
the ring. Since the average value of $\sin^2\alpha$ around the ring is 0.5, we conclude that this total moment is: 
\begin{equation}\label{eq:2.13}
	r^2\sin\theta\frac{GMm}{2R^3}  
\end{equation}
acting about the $y$-axis in the sense to reduce $\theta$.

The other two principal components (\ref{eq:2.8}) are equal, and it will be convenient for later purposes to consider them in combination. We then observe that in combination they will produce a force on a point mass $\delta  m$ which is perpendicular to the $z$-axis, and proportional to the distance of $\delta  m$ from it (and acting away from the $z$-axis, since the sign of (\ref{eq:2.8}) is positive). Since $\cos\theta \approx1$, the distance from the $z$-axis of all point masses on the Moon’s ring is $r$, so from (\ref{eq:2.8}) the force on a single point mass $\delta  m$  is:
\begin{equation}\label{eq:2.14}
	\delta  m \frac{GM}{2R^3}r 
\end{equation} 
The moment this force produces about the $y$-axis, in the sense to reduce $\theta$, is due to its component in the $x$-direction (i.e. (\ref{eq:2.14}) times $\sin\alpha$), and its lever arm about the $y$-axis which is
\begin{equation*}\label{eq:2.15}
	r\sin\alpha \sin\theta
\end{equation*}
see  \mbox{Figure \ref{fig2}}. Thus the moment about the $y$-axis, in the sense to reduce $\theta$, is:
\begin{equation*}\label{eq:2.16}
	\delta  m\frac{GM}{2R^3}r^2 \sin^2\alpha \sin\theta
\end{equation*}
The total moment on the Moon’s ring will again be the sum of the contributions from all the masses $\delta m$ around the ring. We again observe that the average value of $\sin^2\alpha$ around the ring is 0.5, and thus conclude that this total moment is:
\begin{equation}\label{eq:2.17}
 	r^2\sin\theta\frac{GMm}{4R^3}  
\end{equation}
acting about the $y$-axis in the sense to reduce $\theta$.
Thus overall the gradient in the Sun’s gravitational force produces an average moment on the Moon about
$C$ of (\ref{eq:2.13}) + (\ref{eq:2.17}) i.e.
\begin{equation}\label{eq:2.18}
	3r^2\sin\theta\frac{GMm}{4R^3}  \end{equation}
acting about the $y$-axis in the sense to reduce $\theta$.

In vector notation, $\mathbf{S}$ is the unit normal to the plane of the Sun’s orbit [i.e. in the $z$-direction at this stage], and we can define a unit vector $\mathbf{l}$ (for lunar) normal to the plane of the Moon’s orbit, see  \mbox{Figure \ref{fig2}}. The moment can then be written:
\begin{equation}\label{eq:2.19}
	3r^2\frac{GMm}{4R^3}\mathbf{l} \bm{\times} \mathbf{S}.
\end{equation}
Its effect is to make to $\mathbf{l}$ precess about $\mathbf{S}$ with constant angular velocity $\Omega$ (i.e. $\varphi = -\Omega t$ in  \mbox{Figure \ref{fig2}}), as described in  \citet[Section 9.6]{kibble2004classical}. For later use, we will now re-work that argument in a different notation, first writing:
\begin{equation}\label{eq:2.20}
\mathbf{l} = \mathbf{S} \cos\theta + \mathbf{l'} 
\end{equation}
where $\mathbf{l'} $ the component of $\mathbf{l}$  perpendicular to $\mathbf{S}$ (i.e. in the $x$-direction, see  \mbox{Figure \ref{fig2}}). Thus (\ref{eq:2.19}) becomes:
\begin{equation*}\label{eq:2.21}
 	3r^2\frac{GMm}{4R^3}(\mathbf{S}\cos\theta + \mathbf{l'})
	 \bm{\times}
	  \mathbf{S} = 3r^2\frac{GMm}{4R^3} \mathbf{l'} \bm{\times} \mathbf{S} 
\end{equation*}
We can equate this moment about $C$ to the rate-of-change of the Moon’s angular momentum about $C$.  Ignoring the Moon’s much smaller angular momentum about its own CG, the angular momentum of the Moon about $C$ is in the direction of $\mathbf{ l}$ (since the Moon rotates anti-clockwise in  \mbox{Figure \ref{fig2}}) and is:
\begin{equation}\label{eq:2.22}
	 \frac{2\pi  r^2 m}{T} \mathbf{l} =  \frac{2\pi  r^2 m}{T} (\mathbf{S}\cos \theta + \mathbf{l'} )
\end{equation}
 where $T$ is the period of the Moon’s orbit. Thus:
\begin{equation}\label{eq:2.23}
	\frac{d}{dt} 
	\left[
		\frac{2\pi  r^2m}{T}
		(\mathbf{S}\cos\theta + \mathbf{l'})
	\right]
	=
	 3r^2\frac{GMm}{4R^3} \mathbf{l'} \bm{\times}  \mathbf{S} 
\end{equation}
	Since $\mathbf{l'} \bm{\times}  \mathbf{S} $ has no component in the direction of $\mathbf{S}$, we immediately see that \mbox{$\frac{d}{dt}\cos\theta = 0$}, so the
inclination $\theta$ of the Moon’s orbit is a constant. And:
\begin{equation}\label{eq:2.24}
	\frac{d \mathbf{l'}}{dt}
 = \frac{3GMT}{8\pi  R^3}\mathbf{l'} \bm{\times}  \mathbf{S} 
\end{equation}
Although a little excessive in the present case, it will be convenient later to take real and imaginary axes in the $x$-$y$ plane but not rotating relative to distant stars, and hence not aligned with the $x$ and $y$  axes. In this way we can denote the component of a vector in the $x$-$y$ plane by a complex number. The vector $\mathbf{l'}$, in particular, we will denote by the complex number $l'$, where \mbox{$l' = \sin \theta[\cos \varphi + i\sin\varphi]$} in \mbox{Figure \ref{fig2}}. The vector $\mathbf{l'} \bm{\times} \mathbf{ S}$ is also in the $x$-$y$ plane, and in our complex notation it is $-il'$, see  \mbox{Figure \ref{fig2}}. Thus in our complex notation (\ref{eq:2.24}) becomes:
\begin{equation}\label{eq:2.25}
	\frac{dl'}{dt} = -i\frac{3GMT}{8\pi  R^3}l' 
\end{equation}
The solution to this simple differential equation is:
\begin{equation*}\label{eq:2.26}
	l' = a \exp\left(-i\frac{3GMT}{8 \pi  R^3}t \right)
\end{equation*}
where $a$ is a complex constant and $|a| = \sin\theta$, see  \mbox{Figure \ref{fig2}}. Thus, $\mathbf{l'}$ rotates clockwise in  \mbox{Figure \ref{fig2}}, with angular frequency $\Omega = 3GMT/8 \pi  R^3$. The rotation of $\mathbf{l'}$ is the precession of $\mathbf{l}$ described in \citet{kibble2004classical}, where $\Omega$ is derived as equation (9.29), without having recourse to a complex notation. The precession period is  
\begin{equation*}\label{eq:2.27}
	\frac{2\pi }{\Omega}  = \frac{16\pi ^2}{3} \frac{R^3}{GMT}
\end{equation*}
Inserting the known values of $R, G, M$ and $T$ in MKS units (\citet{kibble2004classical} pp. xv, xvi), this comes to:
\begin{equation}\label{eq:2.28}
	\frac{16\pi ^2}{3}
	\frac
		{(1.50 \times10^{11})^3}
		{
				6.67\times10^{-11}\times1.99\times10^{30}\times2.36\times10^6 
		}
	= 5.7\times10^8 \,\text{s} = 18 \,\text{years}
\end{equation}
The agreement with the observed period of 18.6 years \cite{Lunar-precession:2020aa} is reasonable (3 per cent), given our approximations, the most important of which is to ignore the eccentricity of the moon’s orbit, and especially the precession of the direction of the axes of the ellipse (apsidal precession), whose period is 8.85 years \cite{Lunar-precession:2020aa} and therefore long enough to interfere with the precession of the orbit that we are considering.

We have also assumed above that the plane of the Earth’s orbit round the Sun (and thus the plane of the Sun’s orbit as seen in our frame centered on $C$) does not change with time. In fact it is the Laplace invariable plane, whose normal unit vector $\mathbf{L}$ is in the direction of the total angular momentum of the solar system, which does not vary with time. The normal $\mathbf{S}$ to the plane of the Sun’s orbit is inclined at an angle $\xi$ to $\mathbf{L}$ and has a slow precession around it. We will take the $z$-direction in \mbox{Figure \ref{fig2}} as aligned with $\mathbf{L}$, so that $\mathbf{S}$ has a component $\mathbf{S'}$ in the $x$-$y$ plane (now the Laplace invariable plane). Because of the importance of the Moon to the precession of the Earth’s spin axis (see Section \ref{spin}), we need to find the effect of this precession of $\mathbf{S}$, on the precession of $\mathbf{l}$. Instead of (\ref{eq:2.20}), we now have:
\begin{equation}\label{eq:2.29}
	\mathbf{l}= \mathbf{L}\cos\theta+ \mathbf{l'}
\end{equation}
We also now have:
 \begin{equation*}\label{eq:2.30}
	\mathbf{S}= \mathbf{L}\cos\xi + \mathbf{S'}
\end{equation*}
The analysis now follows the development above, except that $\mathbf{l} \bm{\times} \mathbf{ S}$ in (\ref{eq:2.19}) is now:
\begin{equation}\label{eq:2.31}
	[\mathbf{L}\cos\theta + \mathbf{l'}] \bm{\times} [ \mathbf{L}\cos\xi + \mathbf{S'}] 
	= 
	\mathbf{l'} \bm{\times} \mathbf{L}\cos\xi + 
	(\mathbf{L}\cos\theta) \bm{\times} \mathbf{S'} + \mathbf{l'} \bm{\times} \mathbf{ S'}
\end{equation}
Since $|\mathbf{l'}|$ and $|\mathbf{S'}|$ are both $ \ll1$, we can proceed with a first order analysis, in which we ignore their squares and higher powers, and also their product. Thus we can ignore the third term in (\ref{eq:2.31}) and also write $\cos\theta = \cos\xi = 1$, so that (\ref{eq:2.23}) becomes:
\begin{equation}\label{eq:2.32}
 \frac{d}{dt}
 \left[
 	\frac{2\pi  r^2 m}{T}(\mathbf{L} + \mathbf{l'})
\right]
= 
\frac{3r^2 GMm}{4R^3}
(\mathbf{l'} \bm{\times} \mathbf{ L} + 
\mathbf{L} \bm{\times}  \mathbf{S'}) 
\end{equation}
Since $\frac{d\mathbf{ L}}{dt} = 0$ we have:
\begin{equation}\label{eq:2.33}
	\frac{d \mathbf{l'}}{dt} = \frac{3GMT}{8\pi  R^3}(\mathbf{l'} \bm{\times} \mathbf{ L}
  + \mathbf{L} \bm{\times}  \mathbf{S'})
\end{equation}
This equation can be compared with (\ref{eq:2.24}), where $\mathbf{S}$ was assumed fixed and is now replaced by the fixed $\mathbf{L}$, and additionally we have the term $\mathbf{L} \bm{\times} \mathbf{S'}$ on the RHS, which is evidently the effect of the precession of $\mathbf{S}$ about $\mathbf{L}$. We can again replace $3GMT/8\pi  R^3$ with the angular frequency $\Omega$, and now not only denote $\mathbf{l' }$ by the complex number $l'$, but also denote $\mathbf{S' }$ by the complex number $S'$. We now note that $\mathbf{l'} \bm{\times} \mathbf{ L }$ and $\mathbf{L} \bm{\times} \mathbf{S'}$ are denoted by $-il'$ and $iS'$ respectively. Neither our present Earth-centered frame of reference, nor the Sun-centered frame considered later in this paper, rotate with respect to distant stars, and the plane of the Sun’s orbit in the former is clearly the same as the plane of the Earth’s orbit in the latter. The unit normal to this plane is denoted by $\mathbf{S}$ in our present Earth-centered frame, and by $\mathbf{e}$ in the Sun-centered frame. Thus $\mathbf{S} =\mathbf{ e}$ and their complex components $S'$ and $e'$ in the Laplace invariable plane are likewise equal. Thus:
\begin{equation}\label{eq:2.34}
	S' = e' =  b_1 e^{-i\Omega_1 t} + b_2 e^{-i\Omega_2 t} + b_s e^{-i\Omega_3 t}
\end{equation}
where $ \Omega_1,\Omega_2, \Omega_3$ are the angular frequencies, and $b_1, b_2, b_3$ are the complex amplitudes, of the frequency components of the precession of the orbit of the Earth around the Sun, given in Table \ref{table2}. Thus in our complex notation (\ref{eq:2.33}) becomes:
\begin{equation}\label{eq:2.35}
	\frac{dl'}{dt} = -i\Omega(l'-S') = -i\Omega(l' -b_1 e^{-i\Omega_1 t} - b_2 e^{-i\Omega_2 t} - b_s e^{-i\Omega_3 t})
\end{equation}
 This simple differential equation is readily solved as:
 \begin{equation}\label{eq:2.36}
 	l' = ae^{-i\Omega t} +
	 \frac{b_1}{1 -  \Omega_1/\Omega}e^{-i\Omega_1 t} +
	 \frac{b_2}{1 - \Omega_2/\Omega}e^{-i\Omega_2 t} +
	 \frac{b_3}{1 - \Omega_3/\Omega}e^{-i\Omega_3 t} 
 \end{equation}
 where $a$ is again a complex constant. Our interest in Section \ref{spin} will be in the precession of the axis $\mathbf{l}$ of the orbit of the moon relative to the axis $\mathbf{S}$ of the orbit of the Sun, which in our complex notation is:
 \begin{equation}\label{eq:2.37}
 	l' - S' = ae^{-i\Omega t} +
	 \frac{b_1}{\Omega/\Omega_1 -1}e^{-i\Omega_1 t} +
	 \frac{b_2}{ \Omega/\Omega_2 -1}e^{-i\Omega_2 t} +
	 \frac{b_3}{ \Omega/\Omega_3 -1}e^{-i\Omega_3 t}  
 \end{equation}
 
We now observe that $\Omega_1, \Omega_2, \Omega_3  \ll \Omega$ so the last three terms on the RHS of (\ref{eq:2.37}) are very small compared with the first. The axis of Moon’s orbit therefore effectively precesses around $\mathbf{S}$, although $\mathbf{S}$ is itself precessing, much more slowly. This is as observed: the inclination of the Moon’s orbit to that of the Sun is a constant 5.145\text{\textdegree}  throughout the 18-year precession cycle \citet{Williams:ac}. If the precession were not around $\mathbf{S}$, the angle would vary.
\section{Precession of the orbit of the Earth: other planets to be considered}\label{earth}
The precession of the plane of the Earth’s orbit around the Sun is clearly directly relevant to variations of the inclination of the Earth’s spin axis to this plane, which is the main business of this paper. \citet{kibble2004classical}(p.144) explain  that the precession is caused by the gravitational field of the other planets, which can be considered to have their masses smeared into rings along their individual orbits, as we did above for the Sun. This time, we can take our frame of reference as centered on the Sun, and again not rotating relative to distant stars.

The analysis above of the effect of the Sun’s ring on the orbit of the Moon, now applies in exactly the same way to the effect of each of the planets’ rings on the orbit of the Earth. From (\ref{eq:2.18}), the moment applied by each planet to the Earth is proportional to its mass and inversely proportional to the cube of the radius of its orbit (or semi-major axis, the effect of eccentricity being negligible, see Appendix \ref{eccent}). It is necessary to include a factor $f$ which arises because the ratio of the radii of the planetary orbits to that of the Earth (or vice-versa) is often not large, whereas in Section \ref{moon} the radius of the sun’s orbit was very much greater than that of the moon, and $f$ was ignored because it was very close to 1. It is computed in Appendix \ref{orbit}.

The middle column of numbers in Table \ref{table1} gives from \citet{Williams:ab} the figures for $f \times \text{mass}/(\text{radius of orbit})^3 $ for each planet. The figures are shown relative to the figure for Jupiter, which is taken as 1. For planets whose orbit is inside that of the Earth, we observe that the moment they apply to the Earth is equal and opposite to that applied to them by the Earth, which we can evaluate by (\ref{eq:2.18}), including the correction factor $f$. The moment is evidently proportional to the mass of the planet and inversely to the cube of the radius $R$ of the Earth’s orbit, with an additional factor $(r/R)^2$, where $r$ is the radius of the planet’s orbit.
 \begin{table}[htbp]
	\centering 
	\caption{Relative contributions of the planets to the moment on Venus, the Earth and Jupiter (figures $\leq 0.001$ entered as ``Neg.")}\label{table1}
  		\begin{tabular}{lccc}
  			\toprule
			 \multirow{2}{*}{Planet} & \multicolumn{3}{c}{$f \times$ mass/(radius of orbit)$^3$}  \\
	 
			\cmidrule(lr){2-4} \\
		 	{}   & Acting on Venus & Acting on the Earth   & Acting on Jupiter\\
			\midrule
			Mercury & 0.03 & Neg. & Neg.\\
			Venus &- & 0.72 & Neg.\\
			Earth  &1.75&-&Neg.\\
			Mars &0.02 & 0.04 & Neg. \\
			Jupiter &1 & 1 & -\\
			Saturn &0.05 &0.05& 1\\
			Uranus &Neg. & Neg. & 0.01\\
			Neptune &Neg. & Neg. & Neg. \\
			\midrule
			TOTAL & 2.85 & 1.81 & 1.02\\
			\bottomrule
	\end{tabular}
\end{table}

 It can be seen from the middle column of figures that it will be necessary to consider the precession of the orbits of Jupiter and Saturn, to calculate the precession of the orbit of the Earth. The corresponding figures for the planets’ influence on Jupiter are given in the last column of Table 1. It may be seen that the figure for Saturn (now used as the datum instead of Jupiter) is the only significant one. We may therefore consider the precession of the coupled Saturn-Jupiter system in isolation. This is done in Section \ref{jupiter} below, which gives details of the precession of the orbits of both planets.
 
The other significant figure in the middle column is that for Venus, so it will be necessary also to consider the precession of the orbit of Venus. The corresponding figures for the planets’ influence on Venus are given in the remaining column in Table 1. It may be seen that the dominant influence is that of the Earth – we therefore need to consider the precession of the coupled Earth-Venus system. This is done in Section \ref{venus}, and includes the influence of the precession of Jupiter and Saturn, analyzed in Section \ref{jupiter}.
\section{Precession of the orbits of Jupiter and Saturn}\label{jupiter}
The last column of Table \ref{table1} suggests that Jupiter and Saturn can, as a first approximation, be considered in isolation. In that case they constitute the whole angular momentum of the solar system, so the unit normals $\mathbf{j}$ and $\mathbf{s}$ to their orbits must be precessing at the same angular frequency $ \Omega_3$ on either side of $\mathbf{L}$, inclined in inverse proportion to their individual angular momenta. This momenta ratio is given in Table 8 of \textcite{souami2012solar} as $ 61.4/24.9 = 2.46$, so in our complex notation $s' = -2.46j'$. The difference between the arguments of $s'$ and $j'$ is given in Table 9 of \textcite{souami2012solar} as $1.026\pi $ radians, so our approximation appears a fair one. On the other hand the inclinations of their orbits to $\mathbf{L}$ are also given in Table 9, and their ratio is $ 0.925/0.322 = 2.87$ rather than $2.46$, which shows the limitations of the approximation. The reason for the discrepancy is clear from Table 8, which shows that Jupiter and Saturn do not provide the whole angular momentum of the solar system, but only 86 per cent of it, with the remaining 14 per cent coming almost entirely from Uranus and Neptune. Including the precession of these two planets is unwarranted for our present purpose, because we will see in Section \ref{spin} that there is a relatively small contribution from Jupiter and Saturn to the variation of the inclination of the Earth’s spin axis to the plane of its orbit around the Sun. We will therefore simply assume the present-day relationship 
\begin{equation}\label{eq:4.0}
	s' = -2.87j'
\end{equation} and recognize that our approximation is only a rough one. And we will write:
\begin{equation}\label{eq:4.1}
	j' = c e^{-\Omega_{3}t}
\end{equation}
with $c$ denoting the complex amplitude of Jupiter’s precession. It remains to find the angular frequency $\Omega_3$ of the precession. This can easily be done using the methods of the last Section, where we now have Jupiter in place of the Moon, and Saturn in place of the Sun. In place of (\ref{eq:2.33}) we have:
\begin{equation}\label{eq:4.2} 
	\frac{d \mathbf{j'}}{dt} =
	\frac{3G \times 1.95MT}{8 \pi R^3}
	\left[
		\mathbf{j'} \bm{\times} \mathbf{L} - \mathbf{L} \bm{\times} 2.87 \mathbf{j'}
	\right]
\end{equation} 
where $M$ and $R$ are now the mass and orbit radius of Saturn (with the correction factor $f = 1.95$ given in Table \ref{tablea3.1}), and $T$ is the orbital period of Jupiter. In our complex notation this becomes:
\begin{equation}\label{eq:4.3}
	\frac{dj'}{dt} = -i\frac{3G \times 1.95MT}{8 \pi R^3}(1 + 2.87)j'.
\end{equation}
 
This compares directly with (\ref{eq:2.25}). From \textcite{kibble2004classical}(p.xvi) and \citet{Williams:ab} the ratio of the mass of Saturn to that of the Sun is $2.86 \times10^{-4}$, the ratio of the orbit radius of Saturn to that of the Sun is 9.582, and the ratio of the orbital period of Jupiter to that of the moon is 158.5, so scaling from (\ref{eq:2.28}) the precession period $2\pi /\Omega_3$ of Jupiter and Saturn is:
\begin{equation}\label{eq:4.4}
 \frac
 {18 \times 9.582^3}
 {1.95 \times 2.86 \times 10^{-4} \times 158.5 \times 3.87}
  = 50,000 \, 
  \text{years }
\end{equation}
This figure agrees closely with that obtained by \textcite{ stockwell1872memoir}, using the analytical methods of Laplace \cite{laplace1799traite}. He gives (at p.xiv) the long-term average precession rate with great precision, as 25.934567 arcseconds/yr. This rate corresponds to a long-term average precession period of just below 49,972 years.
 
\section{Precession of the orbits of Venus and the Earth}\label{venus}
Using the data in Tables \ref{table1} and \ref{tablea3.1}, we can similarly scale from (\ref{eq:2.28}) to obtain the precession angular frequency of Venus and the Earth, in four alternative scenarios:
\begin{enumerate} [ i)]
	\item Venus precessing under the influence of Jupiter and Saturn alone, whose orbits are assumed fixed (in the Laplace invariable plane). From \textcite{kibble2004classical}(p.xvi) and \citet{ Williams:ab} the ratio of the mass of Jupiter to that of the Sun is $9.54 \times 10^{-4}$, the ratio of the orbit radius of Jupiter to that of the Sun is 5.205, and the ratio of the orbital period of Venus to that of the moon is 8.225. Applying  the factor $f = 1.037$ from Table \ref{tablea3.1}, and the factor 1.047 from Table \ref{table1} to include Saturn, we can scale from (\ref{eq:2.28}) to give the precession angular frequency of Venus (in rad/100,000 yr) in this scenario as:
\begin{equation}\label{eq:5.1}
	\frac
	{(100,000 \times 2\pi /18) \times 9.54 \times 10^{-4} \times 8.225 \times 1.037\times 1.047}
	{5.205^3} 
	= 
	2.11 \,\text{rad/100,000 yr} 
\end{equation}
	\item Venus precessing under the influence of the Earth alone, whose orbit is also assumed fixed (in the Laplace invariable plane). If we remove the factor 1.047 for Saturn, we can scale from (\ref{eq:5.1}) using Table \ref{table1} to give the precession angular frequency of the Earth in this scenario as $(2.11/1.047) \times 1.75 = 3.53$ \mbox{rad/100,000 yr}.
	\item The Earth precessing under the influence of Jupiter and Saturn alone, whose orbits are assumed fixed (in the Laplace invariable plane). From \textcite{kibble2004classical}(p.xvi)  and \citet{ Williams:ab} the ratio of the mass of Jupiter to that of the Sun is $9.54 \times 10^{−4}$, the ratio of the orbit radius of Jupiter to that of the Sun is 5.205, and the ratio of the orbital period of Earth to that of the moon is 13.37. Applying the factor $f = 1.073$ from Table \ref{tablea3.1}, and the factor 1.046 from Table \ref{table1} to include Saturn, we can scale from (\ref{eq:2.28}) to give the precession angular frequency of the Earth (in rad/100,000 yr) in this scenario as:
\begin{equation}\label{eq:5.2}
  	\frac
	{(100,000 \times 2\pi /18) \times 9.54 \times 10^{-4} \times 13.37 \times 1.073\times 1.046}
	{5.205^3}
	 = 3.54 \,\text{rad/100,000 yr} 
\end{equation}
	\item The Earth precessing under the influence of Venus alone, whose orbit is assumed fixed (in the Laplace invariable plane). If we remove the factor 1.046 for Saturn, we can scale from (\ref{eq:5.2}) using Table \ref{table1} to give the precession angular frequency of the Earth in this scenario as $(3.54/1.046) \times 0.72 = 2.44$ \mbox{rad/100,000 yr}.
\end{enumerate}
 
In our complex notation, we can compare with (\ref{eq:2.25}) and see that the differential equations for these four types of precession are respectively:
\begin{align} 
	 \text{i)} \quad \frac{dv'}{dt} &= -i2.11v' \label{eq:5.3}\\
	  \text{ii)} \quad \frac{dv'}{dt} &= -i3.53v' \notag \\
	 \text{iii)} \quad \frac{de'}{dt} &= -i3.54e' \notag \\
	 \text{iv)} \quad \frac{de'}{dt} &= -i2.44e',\notag 
\end{align}	
where the unit of time is in all cases 100,000 years. If we allow Venus and Earth to precess at the same time, with no other planets present, we can compare with (\ref{eq:2.35}) and simply replace $3.53v'$ in (ii) with $3.53(v'- e')$, and $2.44e'$ in (iv) with $2.44(e'- v')$. When we consider the combined scenarios:
\begin{enumerate}[ i)]
\setcounter{enumi}{4}
	\item Venus precessing under the combined influence of the Earth also precessing, and Jupiter and Saturn orbiting without precession in the Laplace invariable plane
	\item The Earth precessing under the combined influence of Venus also precessing, and Jupiter and Saturn orbiting without precession in the Laplace invariable plane
 \end{enumerate}
we can simply add the RHS of the equations (\ref{eq:5.3}) above, to give the differential equations in the new scenarios as respectively:
 \begin{align*}
   	\text{v)}\quad\frac{dv'}{dt} &= -i[2.11v' + 3.53(v'- e')]  \label{eq:5.4}\\
	\text{vi)}\quad\frac{de'}{dt} &= -i[3.54e' + 2.44(e'- v')]  
  \end{align*}

This can conveniently be written as a single vector equation for the coupled Venus-Earth system:
\begin{equation}\label{eq:5.5}
	\frac{d}{dt}
	\begin{bmatrix}
		v'\\
		e'
	\end{bmatrix}
	= -i
	\begin{bmatrix}
		2.11 + 3.53 & -3.53\\
		-2.44 & 3.54 + 2.44
	\end{bmatrix}
	\begin{bmatrix}
		v'\\
		e'
	\end{bmatrix}
\end{equation}
 
The eigenvalues of the matrix are 8.75 and 2.87, and the corresponding eigenvectors are:
\begin{equation*}\label{eq:5.6}
	\begin{bmatrix}
	-1.14\\
	1
	\end{bmatrix}
	,\quad\quad
	\begin{bmatrix}
	1.27\\
	1
	\end{bmatrix}.
\end{equation*} 
Thus the two modes of orbital precession of the Venus-Earth system, including the influence of Jupiter and Saturn both orbiting without precession in the Laplace invariable plane, are:
\begin{enumerate}[ 1.]
	\item The Earth precessing with angular frequency $\Omega_1  = 8.75$ rad/10,000 yr and thus period of $\ 2\pi /8.75 = 72,000$ years. At the same time Venus is precessing at the same angular frequency, with 1.14 times the inclination of the Earth, and the axes of their orbits on opposite sides of $\mathbf{L}$ (i.e. $v' = -1.14e'$).
	\item The Earth precessing with angular frequency $\Omega_2 = 2.87$ rad/10,000 yr and thus period of $ 2\pi /2.87 = 220,000$ years. At the same time Venus is precessing at the same angular frequency, with 1.27 times the inclination of the Earth, and the axes of their orbits on the same side of $\mathbf{L}$ (i.e. $v' = 1.27e'$).
\end{enumerate}
It remains to add in the effect of the 50,000-year period precession of Jupiter and Saturn, described in the previous Section. This can be done in the same way that the precession of the Sun’s orbit was added in (\ref{eq:2.35}). This time we have the precession of both Jupiter and Saturn to consider, and from Table \ref{table1} they contribute respectively the fractions $1/1.047 = 0.955$ and $\ 0.047/1.047= 0.045$ of their combined effect on Venus when both their orbits are fixed, so we should replace $2.11v'$ in (\ref{eq:5.3}(i))  with $2.11(v' - 0.955j' - 0.045s')$. Since we are assuming $s' = -2.87j'$, see Section \ref{jupiter}, so we can write this as $2.11(v' - 0.955j' + 0.045\times2.87j') = 2.11v' - 1.74j'$. The same calculation for the Earth replaces $3.54e'$ in   (\ref{eq:5.3}(iii)) with $3.54(e' - 0.956j' + 0.044\times2.87j') = 3.54e' - 2.94j'$. Thus (\ref{eq:5.5}) becomes:
\begin{equation}\label{eq:5.7}
	\frac{d}{dt}
	\begin{bmatrix}
		v'\\
		e'
	\end{bmatrix}
	= -i
	\begin{bmatrix}
		5.64 & -3.53\\
		-2.44 & 5.98
	\end{bmatrix}
	\begin{bmatrix}
		v'\\
		e'
	\end{bmatrix}
	+
	i
	\begin{bmatrix}
		1.74\\
		2.94
	\end{bmatrix}
	j'
\end{equation} 
It is convenient now to change notation, and no longer consider our variables as $v'$ and $e'$ but instead take our variables as $e'_{v_1}$ and $e'_{v_2}$ , aligned with our two eigenvectors. We have:
\begin{equation}\label{eq:5.8} 
	\begin{bmatrix}
		v'\\
		e'
	\end{bmatrix}
	= 
	\begin{bmatrix}
		-1.14 & 1.27\\
		1 & 1
	\end{bmatrix}
	\begin{bmatrix}
		e'_{v_1}\\
		e'_{v_2}
	\end{bmatrix}
 \text{ and thus }
	 \begin{bmatrix}
		e'_{v_1}\\
		e'_{v_2}
	\end{bmatrix}
	= 
	\begin{bmatrix}
		-0.415 & 0.527\\
		0.415 & 0.473
	\end{bmatrix}
	\begin{bmatrix}
		v'\\
		e'
	\end{bmatrix}
\end{equation}

Thus (\ref{eq:5.7}) becomes:
\begin{align*}
	\frac{d}{dt}
	\begin{bmatrix}
		-1.14 & 1.27 \\
		1 & 1
	\end{bmatrix}
	\begin{bmatrix}
		e'_{v_1}\\
		e'_{v_2}
	\end{bmatrix}
	&= 
	 -i
	\begin{bmatrix}
		5.64 & -3.53\\
		-2.44 & 5.98
	\end{bmatrix}
	\begin{bmatrix}
		-1.14 & 1.27 \\
		1 & 1
	\end{bmatrix}
	\begin{bmatrix}
		e'_{v_1}\\
		e'_{v_2}
	\end{bmatrix}
	+ 
	i
	\begin{bmatrix}
		1.74 \\
		2.94
	\end{bmatrix}
	j' \\[10pt]
	&=  
	 -i
	\begin{bmatrix}
		-9.96 & 3.63\\
		8.76 & 2.88
	\end{bmatrix}
	\begin{bmatrix}
		e'_{v_1}\\
		e'_{v_2}
	\end{bmatrix}
	+
	i
	\begin{bmatrix}
		1.74 \\
		2.94
	\end{bmatrix}
	j'
\end{align*}

i.e.
\begin{align}
	\frac{d}{dt}
	\begin{bmatrix}
		e'_{v_1}\\
		e'_{v_2}
	\end{bmatrix}
	&=
	-i
	\begin{bmatrix}
		-0.415 & 0.527\\
		0.415 & 0.473
	\end{bmatrix}
	\begin{bmatrix}
		-9.96 & 3.63\\
		8.76 & 2.88
	\end{bmatrix}
	\begin{bmatrix}
		e'_{v_1}\\
		e'_{v_2}
	\end{bmatrix}
	+ i
	\begin{bmatrix}
		-0.415 & 0.527\\
		0.415 & 0.473
	\end{bmatrix}
	\begin{bmatrix}
		1.74 \\
		2.94
	\end{bmatrix}
	j'  
	 \notag
	\\[10pt]
	&=
	-i
	\begin{bmatrix}
		8.75 & 0\\
		0 & 2.87
	\end{bmatrix}
	\begin{bmatrix}
		e'_{v_1}\\
		e'_{v_2}
	\end{bmatrix}
	+
	i
	\begin{bmatrix}
		0.827\\
		2.11
	\end{bmatrix}
	j'
	\label{eq:5.10}
\end{align}
 
We can write (\ref{eq:5.10}) as two separate equations:
\begin{align*}
	\frac{de'_{v_1}}{dt}&= -i8.75(e'_{v_1}-0.0945j')= -i8.75(e'_{v_1}-0.0945ce^{-i\Omega_3 t})  \\[10pt]
	\frac{de'_{v_2}}{dt}&= -i2.87(e'_{v_2}-0.735j')= -i2.87(e'_{v_2}-0.735ce^{-i\Omega_3 t})  
\end{align*} 
These are directly comparable with (\ref{eq:2.35}) so our simple solution (\ref{eq:2.36}) applies. Noting also that $ \Omega_3/\Omega_1 =  72,000/50,000 = 1.44 $, and that $ \Omega_3/\Omega_1  =   220,000/50,000 = 4.4$, we thus have:
 \begin{equation}\label{eq:5.13}
 	e'_{v_1}
	=
	b_{v_1}e^{-i8.75t}
	-
	\frac{0.0945}{1 - 1.44}
	ce^{-i12.57t} 	 
	=
	b_{v_1}e^{-i8.75t}
	+
	0.215ce^{-i12.57t},
\end{equation}
and	
\begin{equation}\label{eq:5.14}	
 	e'_{v_2}
	=
	b_{v_2}e^{-i2.87t}
	-
	\frac{0.735}{1 - 4.4}
	ce^{-i12.57t} 
	=
	b_{v_2}e^{-i2.87t}
	+
	0.216ce^{-i12.57t}.
 \end{equation} 
 To calculate the complex amplitudes $b_{v_1}$ and $b_{v_2}$ we return to the original notation using (\ref{eq:5.8}):
 \begin{equation*}\label{eq:5.15}
 	\begin{bmatrix}
		v'\\
		e'
	\end{bmatrix}
	=
	\begin{bmatrix}
		-1.14 & 1.27 \\
		1 & 1
	\end{bmatrix}
	\left\{
		\begin{bmatrix}
			b_{v_1}e^{-i8.75t}\\
			b_{v_2}e^{-i2.87t}
		\end{bmatrix}
		+
		\begin{bmatrix}
			0.215\\
			0.216
		\end{bmatrix}
		ce^{-12.57t}
	\right\}
 \end{equation*}
 i.e.
 \begin{equation}\label{eq:5.16}
 	v' = -1.14b_{v_1}e^{-i8.75t}+ 1.27 b_{v_2}e^{-i2.87t} + 0.029ce^{-i12.57t}
\end{equation} 
and
\begin{equation}\label{eq:5.17}
	e' = b_{v_1}e^{-i8.75t}+ b_{v_2}e^{-i2.87t} + 0.431ce^{-i12.57t} 
 \end{equation}
 We can now substitute the observed values of $v', e'$ and $c$ into (\ref{eq:5.13}) and (\ref{eq:5.14}), and solve for $b_{v_1}$ and $b_{v_2}$.
The observed values of $v', e'$ and $c$ for epoch J2000.0 (which we will take as $t = 0$) are given in Table 9 of \textcite{souami2012solar}. If we take our imaginary axis as zero celestial longitude, then the ascending nodes given in Table 9 are the arguments of $v', e'$and $c$, and the sines of inclinations also given in Table 9 are the moduli of $v', e'$ and $c$. Making these substitutions (\ref{eq:5.16}) and (\ref{eq:5.17}) become:
 \begin{align}
 	0.0376e^{i0.913} - 0.029 \times 0.00562e^{i5.357} &= 0.0376e^{i0.917} = -1.14b_{v_1}+ 1.27 b_{v_2} \label{eq:5.18}\\
	0.0274e^{i4.966} - 0.431 \times 0.00562e^{i5.357} &= 0.0252e^{i4.929} =  b_{v_1}+  b_{v_2} \label{eq:5.19}
 \end{align}
Solving for $b_{v_1}$ and $b_{v_2}$: 
\begin{align*}
	b_{v_1} &= -0.0156e^{i0.917} + 0.0133e^{i4.929} = 0.0262e^{i4.456}  \\
	b_{v_2} &= 0.0156e^{i0.917} + 0.0119e^{i4.929} = 0.0121e^{i0.0631}  
\end{align*}

Inserting these values for $b_{v_1}$ and $b_{v_2}$, and the above value for $c$, into (\ref{eq:5.16}) and (\ref{eq:5.17}) we obtain 
\begin{multline*} 
v' = -1.14 \times 0.0262e^
					{
					i(4.456-8.75t)
					} + 
	1.27 \times 0.0121e^
					{
					i(0.0631 -2.87t)
					} + \\
	0.000163e^
				{
				i(5.357 -12.57t)
				}
\end{multline*}
\begin{multline}\label{eq:5.23}
e' =  0.0262e^
					{
					i(4.456-8.75t)
					} + 
	 0.0121e^
					{
					i(0.0631 -2.87t)
					} + \\
	0.00252e^
				{
				i(5.357 -12.57t)
				}. 
\end{multline}
The final figures are given in tabular form in Table \ref{table2}.

\begin{table}[h] 
	\centering 
	\caption{Frequency components of the precession of the orbits of Venus and the Earth, relative to the Laplace invariable plane (inclination of orbit of Venus 
	$=  \sin^{-1}|a_1 e^{-\Omega_1 t} + a_2 e^{-\Omega_2 t} + a_3 e^{-\Omega_3 t}|$, celestial longitude of ascending node of orbit of Venus 
	$ = \arg(a_1 e^{-\Omega_1 t} + a_2 e^{-\Omega_2 t} + a_3 e^{-\Omega_3 t})$ etc.)}\label{table2}
	\begin{tabular}{c c l l} 
		\toprule
		Angular frequency &  Period (years) & 
		\multicolumn{2}{c}{ Complex amplitudes}
		\\
		\cmidrule{3-4} 
		{} & {} & Venus & The Earth\\
		\midrule
		$\Omega_1$ & 72,000 & $a_1 = 0.0299e^{i1.314}$ & $b_1 = 0.0262e^{i4.456}$\\
			$\Omega_2$ & 220,000 & $a_2 = 0.0154e^{i0.0631}$ & $b_2 = 0.0121e^{i0.0631}$	\\
			$\Omega_3$ & 50,000 & $a_3 = 0.000163e^{i5.357}$ & $b_3 = 0.00252e^{i5.357}$	\\	 
		\midrule
			\multicolumn{2}{c} {Sum} & $0.0375e^{i0.914} $ & $0.0275e^{i1.968}$\\
			
		\bottomrule

	\end{tabular}

\end{table} 
Also shown are the sums $a_1 + a_2 + a_3$ and $b_1 + b_2 + b_3$, which may be seen to check against the observed values given above in (\ref{eq:5.18}) and (\ref{eq:5.19}) to the accuracy of our calculation.

\begin{figure}
	\centering\includegraphics[scale = 0.75]{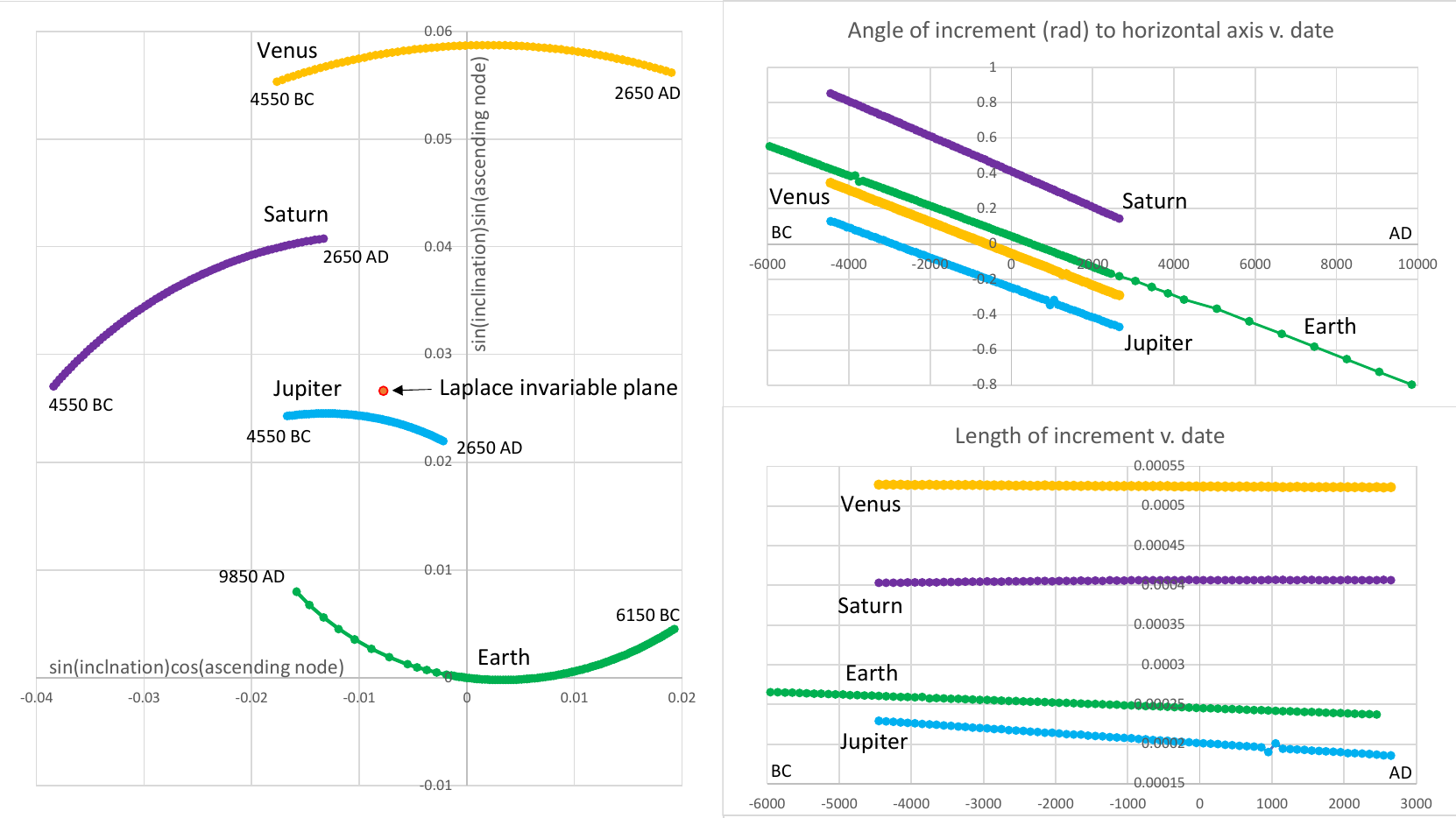}
	\caption{Results from \citet{stockwell1872memoir} for the orbital planes of Venus (Table II), the Earth (Table VIII), Jupiter (Table IV), Saturn (Table V) and the Laplace invariable plane (eqn 531)}\label{figure3}
\end{figure}

\citet{ stockwell1872memoir} gives no figures for the precession period of Venus or the Earth, concluding (p.169) from Laplace’s analytical methods \cite{ laplace1799traite} that the average precession periods of both are indeterminate in the very long term. Stockwell therefore performs a manual integration to find the variation in the earth’s orbit, and the orbits of the other planets, over the last few thousand years. He gives his results as orbital inclinations relative to the ecliptic of 1850, and ascending nodes relative to the March equinox of 1850, every 100 years (sometimes longer). They cover a period of 7,200 years (16,000 in the case of the Earth) and can be plotted in the polar fashion in Figure 6 of \citet{quinn1991three}, as shown on the left in our Figure \mbox{\ref{figure3}}.

The results for the Earth are important in the present context, because they are the data used by Pilgrim to derive his figure of 40,424 years (\citet[ p.61]{pilgrim1904versuch}) for the period of the variations in the tilt of the Earth’s spin axis relative to the plane of its orbit. According to \citet[p.100 ]{imbrie1986ice}, it was Pilgrim’s figure of 40,424 years that was relied on by Milankovitch  in framing his concept of ``Milankovitch cycles". We see in the polar plot in \mbox{Figure \ref{figure3}} that Stockwell’s data appears to form the arc of a circle, like part of the circles for Jupiter and Saturn in \citet[Figure 6]{ quinn1991three}, referred to in the Introduction. The plot for the Earth appears to turn through approximately 1.4 radians in 16,000 years implying an orbital precession period of $16,000\times2\pi /1.4 = 72,000 $years. This is confirmed by the upper graph on the right in \mbox{Figure \ref{figure3}}, which shows the angle of successive increments of the arc to the horizontal axis, as a function of time. It may be seen that angle decreases linearly with time, implying a circular arc, from which more accurate figures for the orbital precession period can be obtained as 73,500 years for the Earth, and 70,500 years for Venus. These figures give some support for the figure of 72,000 years for the main orbital precession mode for both, given in Table \ref{table2}, although it does not support the existence of the other modes in Table \ref{table2}. The figure of 72,000 years becomes a 41,000 year period for the tilt of the earth’s axis to its orbit, when combined with the 26,000-year precession of the Earth’s spin axis (see (\ref{eq:6.2}) – (\ref{eq:6.7}) below) in broad agreement with Pilgrim’s figure of 40,424 years.

Stockwell’s data will not bear closer scrutiny, however. The lower graph on the right in \mbox{Figure \ref{figure3}} shows the length of the successive increments of arc. Those for the Earth and Jupiter decline linearly with time, meaning that the arc of a circle is in fact part of a spiral. In the case of Jupiter, it implies that the orbital inclination will have reduced to zero is less than half an orbital precession period. More seriously, the upper graphs on the right in \mbox{Figure \ref{figure3}} give orbital precession periods of 75,200 years for Jupiter and 63,400 years for Saturn, significantly different from the average figure of just below 49,972 years for both, given elsewhere \cite{ stockwell1872memoir}, see end of previous Section. More seriously still, the orbital inclinations of Jupiter and Saturn can be seen on the polar plot in \mbox{Figure \ref{figure3}} to be on opposite sides of the Laplace invariable plane in 1850 (with their inclinations to it in approximately the ratio 1:3 as they should be, see beginning of previous Section), but this is by no means the case in 4550 BC, when they are both to the left of it. Stockwell’s data for Uranus and Neptune show that in 4550 BC their inclinations too are on the left of the Laplace invariable plane in \mbox{Figure \ref{figure3}} . Since these four planets constitute 99 per cent of the angular momentum of the solar system \cite[Table 8]{souami2012solar}, this angular momentum is not aligned as it should be with the normal to the Laplace invariable plane.

\begin{figure}
	\centering\includegraphics[scale=0.75]{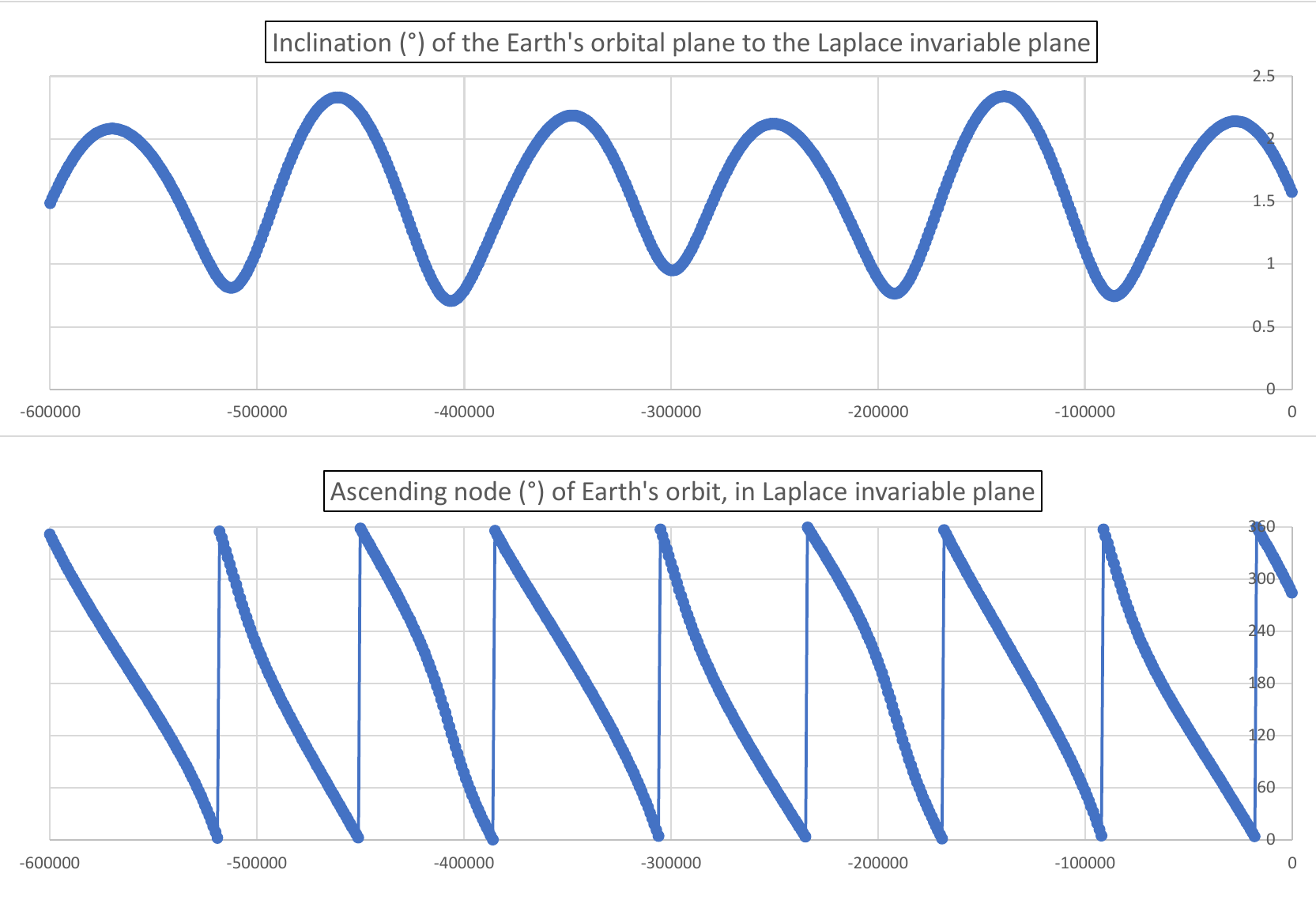}
	\caption{Variation in the inclination and ascending node of the Earth's orbit, relative to the Laplace invariable plane over the past 600,000 years (AD 2000 = 0)}\label{figure4}
\end{figure}

We can also compare our figures for the precession of the Earth’s orbit with the computations of \textcite{ quinn1991three} referred to earlier. That data over the last 600,000 years is replotted in Figure 2.22  of \textcite{muller2002ice} in the form of time-histories of inclinations and ascending nodes referred to the Laplace invariable plane. In our notation these are the inverse sine of the modulus of (\ref{eq:5.23}), and the argument of (\ref{eq:5.23}), respectively. They are plotted in \mbox{Figure \ref{figure4} }, over the same 600,000 year period. 

The agreement with \textcite[Figure 2.22 ]{muller2002ice}, is striking; the graphs have the same two key features highlighted by those authors, namely that the ascending node rotates with period of about 70,000 years, and the inclination varies with a period of about 100,000 years (which they speculate plays a role in the 100,000 year cycle of the ice ages; orbital inclinations to the Laplace invariable plane had apparently not been published before in the paleoclimate literature). Both are consequences of (\ref{eq:5.23}) as we see by factorizing it to: 
\begin{equation*}\label{eq:5.24}
	e' = e^
		{
		i(4.456-8.75t)
		}
			\big[
				0.0262 + 
				0.0121
				e^
					{
						-i(4.393 + 5.88t)
					} + 
				0.00252
				e^
					{
		 				i(0.901-i3.82t)
					}
			\big].
\end{equation*}
Thus the ascending node rotates with an average period of $2\pi /8.75 = 72,000 $ years, as seen in the lower graph of \mbox{Figure \ref{figure4}}, in agreement with both \textcite{muller2002ice} and Stockwell’s data. The phasing is also in agreement with \textcite{muller2002ice}. The rotation is traceable to the first orbital precession mode of the Venus-Earth system, see above.

The inclination is: 
\begin{equation}\label{eq:5.25}
	\sin^{-1}
	| 
		0.0262 + 
				0.0121
				e^
					{
						-i(4.393 + 5.88t)
					} + 
				0.00252
				e^
					{
		 				i(0.901-3.82t)
					}
	|
\end{equation}
The main frequency-component of the variations in inclination thus has a period of $100,000\times2\pi /5.88 = 107,000 $ years, as seen in the upper graph of \mbox{Figure \ref{figure4}}, in agreement with \textcite{muller2002ice}. This variation is not seen in Stockwell’s data. The average magnitude of the variations in inclination in \mbox{Figure \ref{figure4}} agrees reasonably closely with those in Figure 2.22  of \textcite{muller2002ice}, and the phasing of the variations is also in agreement. This main frequency-component is traceable to the second orbital precession mode of the Venus-Earth system, see above. The irregularities in the magnitude of the variations, which are caused by the last term in (\ref{eq:5.25}), traceable to the orbital precession of Jupiter and Saturn, differ somewhat between \mbox{Figure \ref{figure4}} and Figure 2.22  of \textcite{muller2002ice}. This difference may be due to the limitations of our analysis, for example ignoring the influence of Mars, which can be seen from Table \ref{table1} to contribute about 2 per cent to the total influence of the planets on the Earth.

A much more important difference between our results and the computations of Quinn et al. is revealed by their Figure 6, which shows that the inclination of the Earth’s orbit to the Laplace invariable plane is frequently very small. This is not possible with our analysis: according to (\ref{eq:5.25}) its minimum value is 
\begin{equation*}
	\sin^{-1}(0.0262 - 0.0121 - 0.00252) = 0.66\text{\textdegree}.
\end{equation*} It is unclear at this stage whether the difference is due to the limitations of our analysis, or numerical drift in the computations of Quinn et al.
\section{Precession of the spin axis of the Earth, relative to the plane of the Earth’s orbit}\label{spin}
We turn finally the central business of this paper, which is the precession of the spin axis of the Earth, relative to the Earth’s orbit.  \textcite{kibble2004classical} explain (pp 214-215) that it is caused by the gradients in the gravitational field of the Sun and Moon, acting on the slightly ellipsoidal shape of the Earth (\citet[pp 140-143]{kibble2004classical}). They discuss its importance in astronomy, where it is observed as the 26,000-year-period ``precession of the equinoxes", first noticed by the ancient Greeks and analyzed quantitatively by Newton (\citet[ pp 531-534]{ cohen1999principia}). We cannot simply assume that the axis of this precession is fixed, or that it is normal to the plane of the Earth’s orbit. We saw at the end of Section \ref{moon} that it will be somewhere between these two extremes, depending on the ratio of the precession periods.

Our analysis follows Section \ref{moon} closely, with the frame of reference now centered on the CG of the Earth, and the Earth subject to the gradients in the gravitational fields of the Sun and Moon, with the effects of the uniform parts of these fields being cancelled out by the acceleration of the frame. Again the mass of the Sun can be smeared out into a ring, and the Moon too, averaging over the precession of its orbital plane to give a single ring (or not – including its precession allows us to calculate the associated tiny wobble in the precession of the Earth’s axis, described on p.215 of \citet{kibble2004classical} , which gives a useful cross-check on the argument (\ref{eq:2.29}) - (\ref{eq:2.35}). See Appendix \ref{wobble}). It is in the same plane as the Sun’s ring, because the precession axis of the Moon’s orbit is the same as the axis of the Sun’s orbit, to a very close approximation, as shown at the end of Section \ref{moon}.

The components (\ref{eq:2.7}) and (\ref{eq:2.8}) of the gradient of the gravitational field of the Sun are proportional to $ M/R^3$, and can thus be compared with the similar components from the planets in Table \ref{table1}. Since the mass of the Sun is approximately $10^6 $ times that of Jupiter, and it is approximately 5 times closer to Earth, it exceeds the values of  $M/R^3$,in Table \ref{table1} by a factor of more than $10^8$. The mass of the Sun is also greater than that of the Moon by a factor $0.27 \times 10^8$, but the Moon is 389 times closer, so overall its value of  $M/R^3$, exceeds that of the Sun by a factor $\ 389^3/(0.27 \times 10^8) = 2.18$, as noted in p.215 of \citet{kibble2004classical}. Since the gradients of the gravitational fields of the Sun and Moon are both tensors with the same principal axes, we can simply add their components (\ref{eq:2.7}) and (\ref{eq:2.8}).

The next stage of the calculation in Section \ref{moon} is to consider the moment produced by this tensor on a ring which represented the Moon there, but is now one of many rings at various latitudes, forming the out-of-spherical part of the Earth. The moment on each needs to be calculated, and the results added. \mbox{Figure \ref{fig2}} now illustrates a ring at the equator, the other rings will be displaced up or down towards the poles. As the Earth tilts its spin axis to an angle $\psi$ to $\mathbf{S}$, this will introduce from (\ref{eq:2.8}) an additional force in the $x$-$y$ plane on each ring proportional to its latitude and $\sin \psi$. The rings at the same latitude in the northern and southern hemisphere can be paired, so that these forces combine to produce a moment. Since it is also proportional to $\sin \psi$, the combined effect of all the rings is simply another version of (\ref{eq:2.18}), with some other constant rather than $\nicefrac{3}{4}$.

Similarly the angular momentum (\ref{eq:2.22}) will have some other constant representing the Earth’s moment of inertia about its spin axis, and thus the rest of the argument in Section \ref{moon} will be the same, except for different constants. We need not trouble here to calculate them, nor correct them for $\psi$ not being small ($\approx 0.4 $ radians), since the resulting precession period of $2\pi /\Omega_5 = 26,000 $ years given on p.215 of \citet{kibble2004classical} is not at issue – it is the changes in the magnitude $\psi$ of the precession, rather than its period, which is the subject of this paper.

An important observation is that to change the magnitude of $\theta $ in  \mbox{Figure \ref{fig2}}, we require a moment component which is not perpendicular to the angular momentum, i.e. a moment in the $x$-$z$ plane. But if the two rings in  \mbox{Figure \ref{fig2}} are exactly circular, neither an $x$-component or a $z$-component is possible, by symmetry. Thus any changes in the magnitude of $\psi$ must be the result of the orbits of the Sun and Moon not being exactly circular. The Moon’s orbit has a distinct eccentricity (0.055 see \citet{Williams:ac}) but the orientation of the axes of the ellipse is not fixed, it rotates with a period of 8.85 years (apsidal precession \cite{ Lunar-precession:2020aa}). Thus the eccentricity will average out to zero, in a calculation of long-term changes in $\psi$.

The Earth’s orbit has a considerably smaller eccentricity of 0.0167 (see \citet{Williams:2020aa}), but any apsidal precession is much slower, so we must consider this eccentricity more carefully. This is done in Appendix \ref{eccent}, where its effect is shown to be negligible.

All this assumes as at the beginning of Section \ref{moon} that the plane of the Sun’s orbit, with its normal $\mathbf{S}$, is fixed. Thus we conclude that any changes in the inclination $\psi$ of the Earth’s spin axis to $\mathbf{S}$ must be the result of $\mathbf{S}$ not being fixed, but precessing in the way calculated (as the equivalent precession of $e$ around $\mathbf{L}$) in Section \ref{venus} and tabulated in Table \ref{table2}. The effect of the precession of $\mathbf{S}$ on the precession of the orbit of the Moon was calculated at the end of Section \ref{moon}, taking advantage of the fact that the inclination $\theta $ of the axis $\mathbf{l}$ of the Moon’s orbit to $\mathbf{L}$ is small. This time, the equivalent inclination $\chi$ of the unit vector $\mathbf{p}$ in the Earth’s spin axis to $\mathbf{L}$ is about 0.4 radians, so such an approach is questionable. In Appendix \ref{tilt} the analysis below is therefore repeated without assuming that $\chi$ is small, merely that its variations $\delta \chi$ from its average value $\chi_0$ is small. It is shown that the analysis at the end of Section \ref{moon} stands, but with additional correction factors.

The new version of (\ref{eq:2.37}) for the precession of the Earth’s spin axis $\mathbf{p}$ relative to the axis $\mathbf{S}$ of the Sun’s orbit is: 
\begin{equation}\label{eq:6.1}
	p' - S' = de^{-i\Omega_5 t} + 
	0.772 \frac{b_1 e^{-i\Omega_1 t}}{ \Omega_5/\Omega_1 - 1} + 
	0.304 \frac{b_2 e^{-i\Omega_2 t} }{ \Omega_5/\Omega_2 - 1}+
	0.842\frac{ b_3 e^{-i\Omega_3 t} }{ \Omega_5/\Omega_3 - 1} 
\end{equation}
where $p'$ is our complex notation for the component $\mathbf{p'}$ of $\mathbf{p}$ which is perpendicular to $\mathbf{L}$, and $d$ is a complex amplitude. The additional factors 0.772, 0.304 and 0.842 are from Appendix \ref{tilt}, as just described.

Evaluating the fractions numerically using Table \ref{table2} and the value 
\begin{equation*}
	\Omega_5 = 2\pi /0.26\\
	 = 24.17\, \text{ rad/100,000 yr}
\end{equation*}
 given above, we obtain: 
\begin{multline}\label{eq:6.2}
	p' - S' = 0.3970e^{i(3.109-24.17t)} +
	0.0115e^{i(4.456-8.75t)}+\\
	0.000496e^{i(0.0631-2.87t)}+
	0.00230e^{i(5.357-12.57t)}
\end{multline}
where we have also evaluated $d$ numerically, from the fact that $t = 0$ is the year 2000, when the direction of the Sun at the March equinox is zero celestial longitude and our imaginary axis, so $\arg(p' - S') = \pi $ at t = 0. Also $\psi= 0.4091$ radians from \citet{Williams:2020aa}, so $|p' - S'| = \sin 0.4091 = 0.3978$.

\begin{figure}
	\centering\includegraphics[scale=0.75]{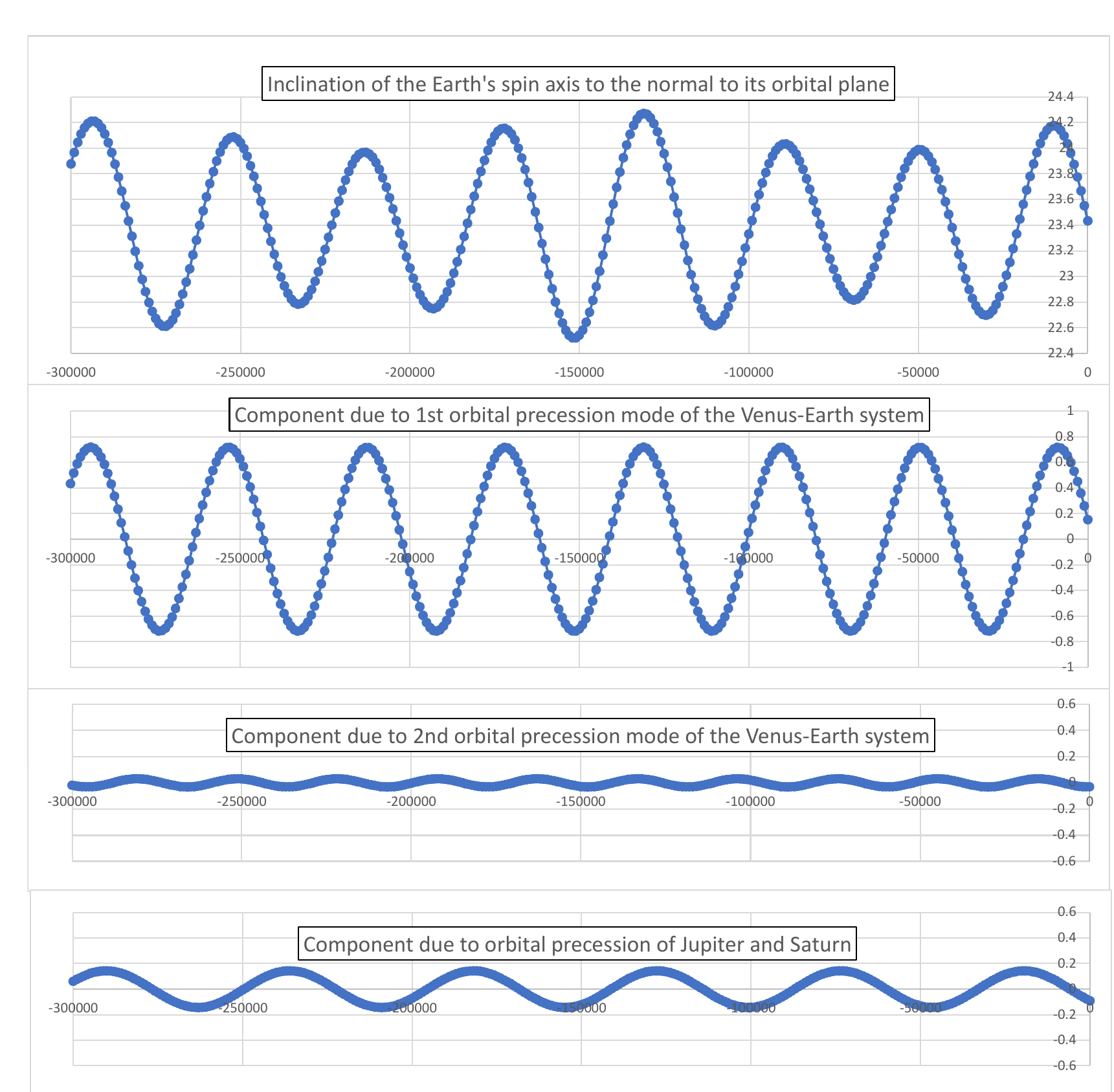}
	\caption{Variation in the inclination in \text{\textdegree} of the Earth's spin axis (relative to the normal to the plane of its orbit) over the past 300,000 years (AD 2000 = 0), with components from different causes}\label{figure5}
\end{figure}

We can factorize (\ref{eq:6.2}) to:
\begin{multline}\label{eq:6.3}
	p' - S' =  e^{i(3.109-24.17t)}(
							0.3970 + 0.0115e^{i(1.347 + 15.42t)} +\\
							0.000496e^{i(3.237 + 21.30t)} +
							0.00230e^{i(2.248 + 11.60t)}
						).
\end{multline}
  The inclination $\psi$ of the Earth’s spin axis to $\mathbf{S}$ is $\sin^{-1}|p' - S'|$, thus:
 \begin{multline}\label{eq:6.4}
 	\psi = \sin^{-1}|  
					0.3970 + 
					0.0115e^{i(1.347 + 15.42t)} + \\
					0.000496e^{i(3.237 + 21.30t)} + 
					0.00230e^{i(2.248 + 11.60t)}
			|
 \end{multline} 
The variation in $\psi$ over the last 300,000 years is shown in \mbox{Figure \ref{figure5}}.

From the form of (\ref{eq:6.4}) we observe that the main frequency-component of the variations in $\psi$ has a period of $100,000\times(2\pi /15.42) = 41,000 $ years, and is caused by the first orbital precession mode of the Venus-Earth system, see previous Section. There are lesser frequency-components of periods $100,000\times(2\pi /21.30)  = 29,500 $ years and $100,000\times(2\pi /11.60)  = 54,000 $ years, caused respectively by the second orbital mode of precession of the Venus-Earth system, and the precession of the orbits of Jupiter and Saturn. To be more precise, if we put $\psi = \psi_0 + \delta \psi$, where $\sin\psi_0 = 0.3970$, and take $\cos(\delta \psi) = 1 $ and $\sin(\delta \psi) = \delta \psi$, then: 
\begin{multline}\label{eq:6.5}
	\sin\psi = \sin\psi_0 + \delta\psi \cos\psi_0
	= |
		0.3970 + 
		0.0115e^{i(1.347 + 15.42t)} + \\
		0.000496e^{i(3.237 + 21.30t)} + 
		0.00230e^{i(2.248 + 11.60t)}
	|
\end{multline}
i.e.
\begin{multline}\label{eq:6.6}
	0.3970 + \delta\psi\cos \psi_0\approx 
	0.3970 +
	0.0115\cos(1.347 + 15.42t) +\\
	0.000496 \cos(3.237 + 21.30t) +
	0.00230\cos(2.248 + 11.60t)
\end{multline}
so that we can obtain the approximate frequency components of $\psi$ as:
\begin{multline}\label{eq:6.7}
	\psi \approx \psi_0 + \frac{1}{{\cos \psi_0}}
	\big[ 
		0.0115\cos (1.347 + 15.42t) + \\
		0.000496 \cos (3.237 + 21.30t) +
		0.00230\cos (2.248 + 11.60t) 
	\big]
\end{multline}
 
These have been added to \mbox{Figure \ref{figure5}}. Note that the relative importance of the second orbital precession mode of the Venus-Earth system (3rd plot), and the orbital precession of Jupiter and Saturn (4th plot) is reversed, compared with their contributions (\ref{eq:5.23}) to the inclination of the Earth’s orbit to the Laplace invariable plane. This is because the different ratios of their periods to the period of the precession of the Earth’s spin axis, and because of the different correction factors to allow for $\chi$ not being small, see (\ref{eq:6.1}).

\mbox{Figure \ref{figure5}} can be compared with the computations of \citet[at their Figure 7(b)]{ quinn1991three}, which is equivalent to the top plot in our \mbox{Figure \ref{figure5}}. The results are similar, with the same 41,000 year period, the same amplitude of about ± 1\textdegree, and the same phasing. \citet{muller2002ice} have carried out further analysis of the results of \citet{ quinn1991three}. They find (p.37) that the additional frequency components of the results of Quinn et al. have periods of 29,000 years and 53,000 years. These are in strikingly close agreement with our figures above. But there are detailed differences, for example the most recent cycle is from 22.7\textdegree to 24.2\textdegree in \mbox{Figure \ref{figure4}}, but from 0.388 to 0.423 radians i.e. from 22.2\textdegree to 24.2\textdegree in Figure 7(b) of \citet{ quinn1991three}. The most significant difference, however, is that the amplitude of the oscillations in Figure 7(b) decays markedly further back in time to 1 million years ago. Since the oscillations are entirely caused by the precession of the Earth’s orbital plane, as we saw above, this is the same discrepancy noted at the end of the previous Section, i.e. that \citet{ quinn1991three} predicts long-term reductions in the magnitude of the orbital precession of the Earth, not seen in our results.

Also shown in Figure 7(b) are the results of the computations of \citet{ berger1978long, berger1976obliquity}, based on the work of \citet{bretagnon1974termes}. This work appears to be a higher-order extension of Lagrange’s perturbation method, and gives results as a series of frequency-components. Up to 100,000 years ago they agree closely with the results of Quinn et. al., but further back in time they start to differ markedly, with Berger’s results, like ours, not showing the long-term reduction in the magnitude of the orbital precession of the Earth, seen in Figure 7(b). Berger’s results do not appear to show any systematic differences from \mbox{Figure \ref{figure5}}: although the most recent cycle is larger than that shown in Figure 2b of \citet{ berger1976obliquity}, by the fourth cycle the position is reversed.

Overall, the broad agreement between all three sets of results very strongly suggest that all three are essentially correct, since they were obtained by three quite different methods. They also broadly agree with two of the results (from \cite{ laskar2004long, Sharaf:1969aa}) in  \mbox{Figure \ref{fig1}}. This agreement very strongly suggests that the third (from \citet{ smulsky2016fundamental}) is incorrect. Further light is shed by Figure 15 of \citet{laskar2004long} which gives the variation in the Earth’s obliquity over the last million years, in exactly the same format as Figure 7(b). The agreement is extremely impressive, given the completely different methods employed. This strongly suggests that the long-term reduction in the magnitude of the orbital precession of the Earth, seen in both, is correct. The discrepancy with our results and Berger’s, on this point, is evidently caused by the limitations of our analyses.
\section{Acknowlegement}\label{ack}
I am grateful to Dr. A. P. Hamblin of the Geological Survey of Canada, for drawing my attention to this problem. Also to Mr. J. G. Colman of the School of Mathematics and Statistics of the University of Sheffield, for detailed reading of successive drafts of the manuscript, leading to numerous important clarifications. I would like to thank Professor S. D. Tremaine of the Institute for Advanced Study, Princeton for pointing out the agreement between \citet{laskar2004long} and \citet{quinn1991three} just noted.

\section{References}
\bibliographystyle{rss}
\bibliography{rctr} 
 
\appendix
\numberwithin{equation}{section}
\numberwithin{figure}{section}
\numberwithin{table}{section}
\section{ Notation}\label{app1}
Although symbols are defined as necessary when they are first introduced, a tabular presentation of the main symbols may also be helpful, see Tables \ref{tablea1.1a} and \ref{tablea1.1b} (where $\mathbf{L}$ is a unit normal to the Laplace invariable plane: $\mathbf{L}$ is aligned with the total angular momentum of the solar system, and thus does not vary with time).
 \begin{table}[htbp]
	\centering  
	\caption{Unit normals and their components}\label{tablea1.1a}
	\renewcommand{\tabularxcolumn}[1]{m{#1}}
		\begin{tabularx}{0.8\textwidth}{l l >{\centering\arraybackslash}Xcc}
		\toprule
		 	 {Body} & 
			 {Orbiting}& 
			{Unit normal along axis of orbit}& 
			\multicolumn{2}{c}{Component normal to $\mathbf{L}$ } \\
	 
			\cmidrule(l){4-5} \\
		 	{}   & {}&
			{}&
			\text{Vector notation} & \text{Complex notation}  \\
 
			\midrule 
			Moon&
			Earth& 
			 $\mathbf{l}$ &
			 $\mathbf{l'}$ &
			 $l'$\\
			 
			 Sun&
			 Earth& 
			 $\mathbf{S}$ &
			 $\mathbf{S'}$ &
			 $S'$\\
			 Venus&
			 Sun&
			 $\mathbf{v}$ &
			 $\mathbf{v'}$ &
			 $v'$\\
			 Earth&
			 Sun&
			 $\mathbf{e}$ &
			 $\mathbf{e'}$ &
			 $e'$\\
			 Jupiter&Sun&
			 $\mathbf{j}$ &
			 $\mathbf{j'}$ &
			 $j'$\\       
			 Saturn&
			 Sun&                                                                                                                                                                                                                                                                                               
			 $\mathbf{s}$ &
			 $\mathbf{s'}$ &
			 $s'$\\
			 Earth&
			 its axis&
			 $\mathbf{p}$& 
			  $\mathbf{p'}$ & 
			 $p'$ \\
						\bottomrule
	\end{tabularx}
\end{table}

   \begin{table}[htbp]
	\centering  
	\caption{ }\label{tablea1.1b} 
		\begin{tabularx}{ 0.91\textwidth}{l l ccc}
		\toprule
			{Body} & 
			{Orbiting}&
			 Frequency components&
			 Ref.&
			Inclinations 			 
		\\	
		
		\midrule 
			Moon&
			Earth& 
			$ l' = a e^{-i\Omega t} + ...$&
			(\ref{eq:2.36})&
			$\theta = \sin^{-1}|l'|$
		\\
			Sun&
			 Earth& 
			 $ S' = b_1 e^{-i\Omega_1 t}
			 +b_2 e^{-i\Omega_2 t}
			 +b_3 e^{-i\Omega_3 t}			 
			 $&
			 (\ref{eq:2.34})&
			$\xi = \sin^{-1}|S'|$
			\\
			 Venus&
			 Sun&
			 $ v' = a_1 e^{-i\Omega_1 t}
			 +a_2 e^{-i\Omega_2 t}
			 +a_3 e^{-i\Omega_3 t}
			 $&
			 Table \ref{table2}&
			{}
			\\
			 Earth&
			 Sun&
			 $e'= b_1 e^{-i\Omega_1 t}
			 +b_2 e^{-i\Omega_2 t}
			 +b_3 e^{-i\Omega_3 t}			 
			 $&
			 Table \ref{table2}&
			$\xi = \sin^{-1}|e'|$
			\\
			 Jupiter&
			 Sun&
			 $ j' = c e^{-i\Omega_3 t}$&
			 (\ref{eq:4.1})&
			{}
			\\       
			 Saturn & 
			 Sun&  
			 $ s' = -2.87c e^{-i\Omega_3 t}$&
			 (\ref{eq:4.0})&
			{}
		 \\ 

			\multirow{2}{*}{Earth} & 
			\multirow{2}{*}{its axis}&  
			 \multirow{2}{*}{$ p' = d e^{-i\Omega_5 t} + ...$}&
			 \multirow{2}{*}{(\ref{eq:6.1})}&
	 		 $
				\chi = \sin^{-1}|p'|
			$ 
		\\
		 	{}   & {}&
			{}&
			{}&
			$
			(\psi = \sin^{-1}|p'-S'| )
			$ \\ 
			\bottomrule
	\end{tabularx}
\end{table} 
 \section{ Effect of orbit eccentricity}\label{eccent}
\begin{figure}
	\centering\includegraphics{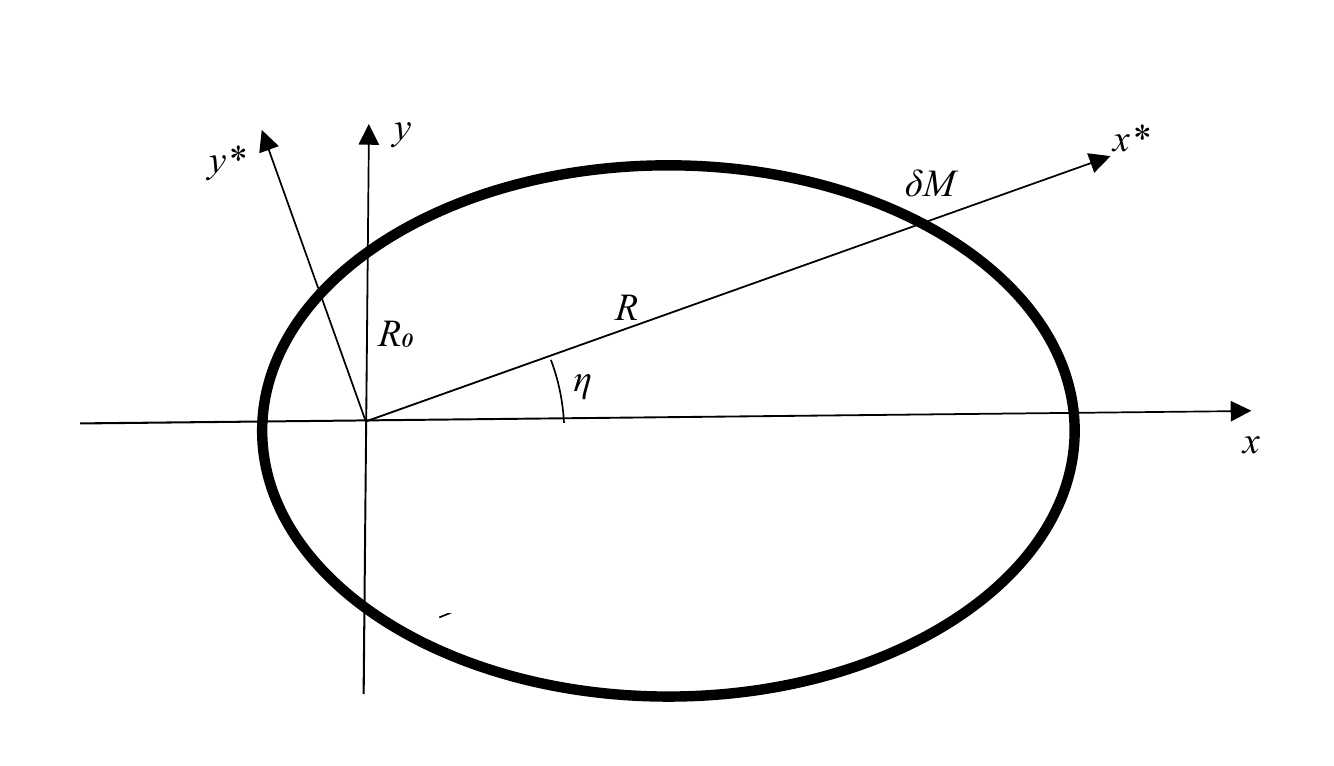}
	\caption{Elliptical orbit of a planet around the Sun}\label{figure2.1}
\end{figure}

The eccentricity of a planetary orbit will affect the calculation in Section \ref{moon} of the tensor gradient of the gravitational field produced when its mass is smeared out over its orbit. \mbox{Figure \ref{figure2.1}} shows this eccentricity in exaggerated form, with the mass of the planet now smeared out over its orbit to form an elliptical ring. We are now taking the origin of coordinates at a focus of the ellipse, and will align the $x$-axis with the major axis of the ellipse.

 The tensor gradient of the planet’s gravitational field will be the sum of tensor gradients from each part of the ring. A single increment $ \delta M $ of the planet’s mass $M$ smeared over the ring, at a distance $R$ from the focus, has a tensor gradient at the focus which can be expressed as a matrix:
\begin{equation}\label{eq:A2.1}
	\frac{G\delta M}{R^3}
	\begin{bmatrix}
		2 & 0 & 0 \\
		0 & -1 & 0 \\
		0 & 0 & -1 \\
	\end{bmatrix}
\end{equation}
where the first coordinate $x^*$ is in the direction of the point mass and at an angle $\eta$ to the $x$-axis, the second coordinate $y^*$ is perpendicular to the $x^*$-axis, and the $z^*$-axis is perpendicular to both. See \mbox{Figure \ref{figure2.1}}. In the $x$-$y$-$z$ coordinate system, this matrix becomes:
\begin{equation}\label{eq:A2.2}
	\frac{G\delta M}{R^3}
	\bm{\Psi}
	\begin{bmatrix}
		2 & 0 & 0 \\
		0 & -1 & 0 \\
		0 & 0 & -1 \\
	\end{bmatrix}
	\bm{\Psi}^{T}
	=
	\frac{G\delta M}{R^3}
	\begin{bmatrix}
		2 - 3 \sin^2 \eta & 3\sin \eta \cos \eta& 0 \\
		3\sin \eta \cos \eta&  2 - 3 \cos^2 \eta & 0 \\
		0 & 0 & -1 \\
	\end{bmatrix}
\end{equation}
where
\begin{equation*}
	\bm{\Psi}  = 
	\begin{bmatrix}
		\cos \eta & -\sin \eta & 0 \\
		\sin \eta &  \cos\eta & 0 \\
		0 & 0 & 1 \\
	\end{bmatrix}
\end{equation*}
The planet’s ring is elliptical, with $ R_0/R  = 1 - \epsilon \cos \eta $ where $R_0$ is the semi-latus rectum (see \mbox{Figure \ref{figure2.1}}) and $\epsilon$ is the eccentricity. By Kepler’s second law (\citet{kibble2004classical} p. 57), the planet’s angular velocity $\frac{d \eta}{dt}$  is proportional to $R^{-2}$, so the length of time $\delta t$ required for an increment $\delta \eta$ of $\eta$, is proportional to $R^2\delta\eta$. The planet’s mass is distributed around the ring in proportion to the time spent there, so $\delta M$ is also proportional to $R^2\delta\eta$. In one complete orbit, the increments $\delta M$ add up to $M$, and the increments $R^2\delta\eta$ add up to twice the area of the orbit i.e. to $ 2\pi R_0^2/(1 - \epsilon^2)$, so we conclude:
\begin{equation}\label{eq:A2.3}
	\delta M = \frac
				{MR^{2}(1 - \epsilon^2)}
				{2\pi R_0^2}
			\delta \eta
\end{equation} 
 We can thus substitute into (\ref{eq:A2.2}) to obtain the tensor gradient from $\delta M$, in the $x$-$y$-$z$ coordinate system, as
the matrix: 
\begin{equation}\label{eq:A2.4}
	\frac
	{GM(1 - \epsilon^2)}
	{2\pi R_0^3}
	\{1 - \epsilon \cos \eta\}\delta\eta
	\begin{bmatrix}
		2 - 3 \sin^2 \eta & 3\sin \eta \cos \eta& 0 \\
		3\sin \eta \cos \eta&  2 - 3 \cos^2 \eta & 0 \\
		0 & 0 & -1 \\
	\end{bmatrix}
\end{equation}
 The tensor gradient from the whole ring is obtained by integrating over $0 < \eta < 2\pi$. We can split (\ref{eq:A2.4}) into two parts corresponding to the two terms in the $\{\}$. The second part integrates to zero, for every element in the matrix. The first part integrates to:
 \begin{equation}\label{eq:A2.5}
 	\frac
	{GM(1 - \epsilon^2) }
	{ R_0^3}
	\begin{bmatrix}
		\frac{1}{2} & 0 & 0 \\
		0 & \frac{1}{2} & 0\\
		0 & 0 & -1
	\end{bmatrix}
	=
	\frac
	{GM  }
	{ R_a^3(1 - \epsilon^2)^2}
	\begin{bmatrix}
		\frac{1}{2} & 0 & 0 \\
		0 & \frac{1}{2} & 0\\
		0 & 0 & -1
	\end{bmatrix}
 \end{equation}
When $\epsilon = 0$ the LHS is recognizable as the tensor components in (\ref{eq:2.7}) and (\ref{eq:2.8}), which were calculated much more simply by symmetry. The tensor retains the symmetry it has for a circular orbit, when the orbit is elliptical. If we use the semi-major axis $R_a$ in place of $R_0$, as on the RHS, we see that the only effect of the eccentricity on the tensor is to increase its overall magnitude by a factor of $(1 - \epsilon^2)^{-2}$. In the case of the Jupiter’s orbit, for example, the eccentricity $\epsilon$ is 0.0489, see \citet{ Williams:ab}, so the factor is 1.005, which is negligible.
\section{ Effect of finite ratios of orbit sizes}\label{orbit}
\begin{figure}
	\centering\includegraphics{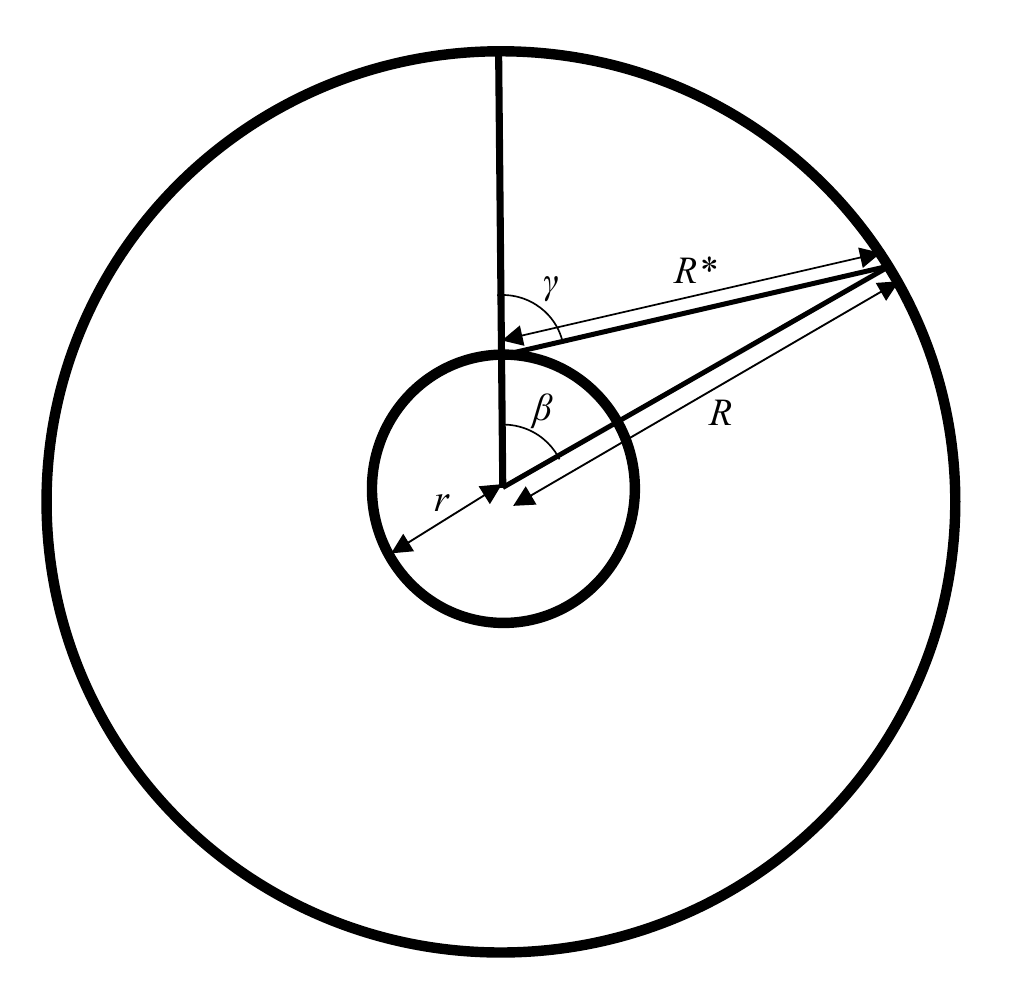}
	\caption{Concentric orbits, e.g. of the Sun and Moon as in Figure 2}\label{figure3.1}
\end{figure}

In Section \ref{moon}, we assumed the (tensor) gradient of the Sun’s gravitational field was constant over the area of the Moon’s orbit. We now take account of the finite ratio $r/R$ of the radii of the orbits of the Moon and Sun. In (\ref{eq:2.13}) we used the principal component (\ref{eq:2.7}) of the gradient in the direction normal to the Sun’s orbit, evaluated at the center of the Moon’s orbit. More accurately, it can be evaluated at each point on the mean position of the Moon’s orbit – it has the same value at each point, by symmetry. Instead of (\ref{eq:2.7})  it is:
\begin{equation}\label{eq:A3.1}
	\frac{-GM}{2\pi}\int_0^{2\pi}\frac{d\beta}{R^{*3}}
	=
	\frac{-GM}{2\pi}\int_0^{2\pi}\frac{d\beta}{(R^2 + r^2 - 2Rr\cos \beta)^{3/2} }
\end{equation}
 see \mbox{Figure \ref{figure3.1}}. This expression can be written:
 \begin{equation}\label{eq:A3.2}
 	\frac{-GM}{R^3} 
	\left\{
			\frac{1}{2\pi}
			\bigintsss_0^{2\pi}
			\frac{d\beta}{
				\left[
					1 + (\frac{r}{R})^2-2\frac{r}{R}\cos \beta
				\right]^{3/2}
			} 
	\right\}
 \end{equation}   
which is (\ref{eq:2.7}) times the multiplier in the $\{\}$ that can readily be evaluated numerically.
 
In (\ref{eq:2.17}) we used the other principal components (\ref{eq:2.8}) of the gradient, again evaluated at the centre of the Moon’s orbit, to calculate the radial force on each point of the Moon’s ring, at its mean position. More accurately, the radial force can be calculated directly – it has the same value at each point, by symmetry. Instead of (\ref{eq:2.14}) it is: 
 \begin{equation}\label{eq:A3.3}
 \delta m
	\frac{GM}{2\pi}
	\int_0^{2\pi}
	\frac{\cos \gamma}{R^{*2}}
	d\beta
	=
	\delta m
	\frac{GM}{2\pi}
	\bigintss_0^{2\pi}
	\frac{\sqrt{1 - 
			\frac
				{R^2 \sin^2 \beta}
				{R^2 +r^2 - 2Rr\cos \beta}
			}}
		{R^2 +r^2 - 2Rr\cos \beta}
		d\beta
\end{equation}
see \mbox{Figure \ref{figure3.1}}, where $R^*\sin \gamma = R\sin \beta$ so that $\cos \gamma = \sqrt{1 - \frac{R^2\sin^2\beta}{R^{*2}}}$, with due attention to sign. The expression \ref{eq:A3.3} can be written:
\begin{equation} \label{eq:A3.4}
	\delta m\frac{GM}{2R^3}r
	\left\{
		\frac{R}{\pi r}
		\bigints_0^{2\pi} 
		\frac
		{\sqrt{
				\left[ 
					1 - 2 \frac{r}{R}\cos \beta + 
						\left(
						\frac{r}{R}
						\right)^2 
				\right] -  
				\sin^2 \beta 
			}
		}
		{
			\left[
				1 - 2 \frac{r}{R}\cos \beta + \left(\frac{r}{R}\right)^2 
			\right]^{3/2}
		}
		\,
		d\beta	
	\right\}	
\end{equation}
which is (\ref{eq:2.14}) times the multiplier in the $\{\}$ that can readily be evaluated numerically.

\begin{table}[htbp]
	\centering 
	\caption{Multipliers to account for finite ratios of orbit sizes}\label{tablea3.1}
  		\begin{tabular}{lccc}
  			\toprule
			 \multirow{2}{*}{Planet} & \multicolumn{3}{c}{Multiplier on overall moment (\ref{eq:2.18})} \\
	 
			\cmidrule(l){2-4} \\
		 	{}   & \text{Acting on Venus} & \text{Acting on the Earth}   & \text{Acting on Jupiter}\\
			\midrule
			Mercury & 1.89 & 1.36 & 1.010\\
			Venus &- & 4.09 & 1.037\\
			Earth  &4.09&-& 1.073\\
			Mars &1.62 & 2.91 & 1.19 \\
			Jupiter &1.037 & 1.073 & -\\
			Saturn &1.011 &1.021& 1.95\\
			Uranus &1.0027 & 1.0051 & 1.15\\
			Neptune &1.0011 & 1.0021 & 1.059 \\
			\bottomrule
	\end{tabular}
\end{table}

The multiplier on the overall moment (\ref{eq:2.18}), is the two multipliers from (\ref{eq:A3.2}) and (\ref{eq:A3.4}) combined, weighted in the ratio 2:1, to reflect the relative contributions of (\ref{eq:2.7}) and (\ref{eq:2.8}) to the overall moment. It is given in Table \ref{tablea3.1} for all the planets as they act on Venus, the Earth and Jupiter respectively.
\section{Wobble in Earth’s spin axis produced by the precession of the Moon’s orbit}\label{wobble}
 \citet{kibble2004classical} note on p.215 that the precession of the orbit of the moon, described in Section \ref{moon}, leads to a wobble of the Earth’s spin axis of the same period, and amplitude 9$''$ = 0.0025\textdegree. The amplitude can easily be calculated by the methods of Sections \ref{moon} and \ref{spin}, and gives a useful cross-check on them. In Section \ref{spin}, the precession of the orbit of the moon adds via (\ref{eq:2.36}) an additional term to the RHS of (\ref{eq:6.1}) so that it becomes:
\begin{multline}\label{A4.1}
	p' - S' = de^{-i\Omega_5 t} + 
	\frac{ b_1}{\Omega_5/\Omega_1 - 1}e^{-i\Omega_1 t} + 	
	\frac{b_2}{ \Omega_5/\Omega_2 - 1}e^{-i\Omega_2 t} +\\
	\frac{b_3}{ \Omega_5/\Omega_3 - 1}e^{-i\Omega_3 t}  +
	\frac{2.18}{2.18 + 1}
	\frac{ae^{-i\Omega t}}{1-  \Omega/\Omega_5}
\end{multline}
the factor $ 2.18/(2.18+1)$ being the fraction of the total gravitational field gradient coming from the Moon, see Section \ref{spin}.

Since $ \Omega/\Omega_5 =  26,000/18$ and $|a| = 5.145$\text{\textdegree} see \citet{Williams:ac}, the additional precession in the Earth’s spin axis has a period of 18 years and an amplitude $(2.18/3.18)\times 5.145 \times (18/26000) = 0.0025$\text{\textdegree}, as observed.
\section{Effect of the tilt of the Earth’s spin axis not being small}\label{tilt}
We now find the effect of the precession of $\mathbf{S}$ on the precession of the Earth’s spin axis $\mathbf{p}$, allowing for the fact that inclination $\chi$ of $\mathbf{p}$ to $\mathbf{L}$ is not small ($\approx 0.4 $ radians). 
We will write $\chi = \chi_0+ \delta \chi$  and $\mathbf{p'}= \mathbf{p'_0}+ \delta \mathbf{p'}$ where $\chi_0$ and $\mathbf{p'_0}$ are the values of $\chi$ and $\mathbf{p}$ calculated ignoring the precession of $\mathbf{S}$ (so that $\chi_0$ and $|\mathbf{p'_0}|$) are constants, with $\sin\chi_0 = |\mathbf{p'_0}|$. We will take $\delta \chi$ and $|\mathbf{\delta p'}|$ to be small, but not $\chi_0$ and $|\mathbf{p'_0}|$. We first have:
\begin{align*} 
	|\mathbf{p'_0}| + \frac
					{\mathbf{\delta p'.p'_0}}
					{|\mathbf{p'_0}|} 
	&= 
	\sin(\chi_0 + \delta \chi )\\
	&= 
	\sin\chi_0 + \delta \chi\cos\chi_0  \\
	&= 
	|\mathbf{p'_0}| + \delta \chi\cos\chi_0  
\end{align*}
so that: 
\begin{equation*} 
	\delta \chi = 
	\frac
		{\delta \mathbf{p'.p'_0}}
		{|\mathbf{p'_0}| \cos\chi_0} 
\end{equation*}
Our analysis in Section \ref{spin} begins with (\ref{eq:6.1}), which is derived from the equivalent analysis of the precession of the Moon’s orbit in Section \ref{moon}, with the normal $\mathbf{l}$ to the Moon’s orbit replaced by $\mathbf{p}$. We now need to replace $\mathbf{L}\cos\theta + \mathbf{l'}$ in (\ref{eq:2.29}) with $\mathbf{L}\cos(\chi_0 + \delta \chi) + \mathbf{p'_0 + \delta p'}$. We have:
 \begin{align*} 
 	\cos(\chi_0 + \delta \chi) 
	&= 
	\cos\chi_0 - \delta \chi\sin\chi_0 \\
	&= \cos\chi_0 - 
		\frac
			{\mathbf{\delta p'.p'_0}}
			{|\mathbf{p'_0}| \cos\chi_0}
		 \sin\chi_0 \\
	&= \cos \chi_0 - 
		\frac
			{\mathbf{\delta p'.p'_0 ́}}
			{|\mathbf{p'_0}|}
		\tan\chi_0
\end{align*}
The analysis now follows that in Section \ref{moon}, with (\ref{eq:2.31}) now becoming: 
 \begin{multline*} 
	\left[
		\mathbf{L}
			\left(
				\cos\chi_0 -
				 \tan\chi_0
				\frac
					{\boldsymbol{\delta}\mathbf{  p'.p'_0}}
					{|\mathbf{p'_0}|}
			\right) + 
			\mathbf{p'_0} + \boldsymbol{\delta}\mathbf{ p'}
	\right] 
	\bm{\times} 
	[ 
		\mathbf{L + S'}
	]
	= \\
	\mathbf{p'_0}\bm{\times}\mathbf{L} + 
	\boldsymbol{\delta} \mathbf{p'}\bm{\times}\mathbf{L} + 
	\left(
				\cos\chi_0 -
				 \tan\chi_0
				\frac
					{\boldsymbol{\delta}\mathbf{  p'.p'_0}}
					{|\mathbf{p'_0}|}
			\right) \mathbf{L}\bm{\times} \mathbf{S'} + 
			\mathbf{p'_0}\bm{\times} \mathbf{S'}  +
			\delta\mathbf{p'} \bm{\times} \mathbf{S'}
\end{multline*}
Ignoring products of small quantities as in Section \ref{moon} this expression becomes:
 \begin{equation*} 
  	\mathbf{p'_0}\bm{\times}\mathbf{L} + 
	\boldsymbol{\delta} \mathbf{p'}\bm{\times}\mathbf{L} +  
	\cos\chi_0 ( \mathbf{L} \bm{\times} \mathbf{S'} )+ 
	\mathbf{p'_0}\bm{\times} \mathbf{S'} 
 \end{equation*}
Thus (\ref{eq:2.32}) becomes:
\begin{multline*} 
	\frac{d}{dt}
	\left[
		\mathbf{L}
			\left(
				\cos\chi_0 -
				\frac
					{\boldsymbol{\delta}\mathbf{  p'.p'_0}}
					{|\mathbf{p'_0}|}
				 \tan\chi_0
			\right) + 
			\mathbf{p'_0} + \boldsymbol{\delta}\mathbf{ p'}
	\right]
	=\\
	\Omega_5
	\left[
		\mathbf{p'_0}\bm{\times}\mathbf{L} +  
		\boldsymbol{\delta} \mathbf{p'}\bm{\times}\mathbf{L} +   
		\cos\chi_0(\mathbf{L}  \bm{\times} \mathbf{S'} )+  
		\mathbf{p'_0}\bm{\times} \mathbf{S'} 
	\right]
\end{multline*}
Since $\frac{d\mathbf{L}}{dt} = 0$ and $\frac{d\mathbf{p'_0}}{dt} = \Omega_5(\mathbf{p'_0} \bm{\times} \mathbf{L})$, this equation reduces to:
\begin{equation*} 
	-\mathbf{L} \tan \chi_0 
	\frac{d}{dt}
	\frac 
		{\boldsymbol{\delta}\mathbf{  p'.p'_0}}
		{|\mathbf{p'_0}|} +
	\frac
		{d}
		{dt}
		(\boldsymbol{\delta}\mathbf{  p'})
	=
	\Omega_5
	\left[ 
		\boldsymbol{\delta} \mathbf{p'}\bm{\times}\mathbf{L} +   
		\cos\chi_0(\mathbf{L}  \bm{\times} \mathbf{S'})+  
		\mathbf{p'_0}\bm{\times} \mathbf{S'}  
	\right]
\end{equation*}
Equating components perpendicular to $\mathbf{L}$:
 \begin{equation*} 
	\frac
		{d}
		{dt}
		(\boldsymbol{\delta}\mathbf{  p'})
	=
	\Omega_5
	\left[ 
		\boldsymbol{\delta} \mathbf{p'}\bm{\times}\mathbf{L} +   
		\cos\chi_0(\mathbf{L}  \bm{\times} \mathbf{S'}  )	\right]
\end{equation*}
This is very similar to (\ref{eq:2.33}) so we can follow the argument (\ref{eq:2.33}) – (\ref{eq:2.36}) in Section \ref{moon} to obtain the complex $\delta p'$ as:
\begin{equation*} 
	\delta p' =
	\delta de^{-i\Omega_5 t} +
	\frac
		{b_1 \cos \chi_0}
		{1-  \Omega_1/\Omega_5}
	e^{-i\Omega_1 t} +
	\frac
		{b_2 \cos \chi_0}
		{1-  \Omega_2/\Omega_5}
	e^{-i\Omega_2 t} +
	\frac
		{b_3 \cos \chi_0}
		{1-  \Omega_3/\Omega_5}
	e^{-i\Omega_3 t} 
\end{equation*}
where $\delta d $ is a complex amplitude. Since 
$\frac
	{d\mathbf{p'_0}}
	{dt} 
= 
\Omega_5(\mathbf{p'_0} \bm{\times} \mathbf{L}) $,
we have 
$p'_0= de^{-i\Omega_5 t}$
where $d$ is another complex amplitude so that:
\begin{multline*} 
	p' = p'_0 + \delta p' = 
	\\(d + \delta d)e^{-i\Omega_5 t} +
	\frac
		{b_1 \cos \chi_0}
		{1-  \Omega_1/\Omega_5}
	e^{-i\Omega_1 t} +
	\frac
		{b_2 \cos \chi_0}
		{1-  \Omega_2/\Omega_5}
	e^{-i\Omega_2 t} +
	\frac
		{b_3 \cos \chi_0}
		{1-  \Omega_3/\Omega_5}
	e^{-i\Omega_3 t}
\end{multline*}
We can thus set $\delta d = 0$ by suitable choice of $d$. Following the argument on to (\ref{eq:2.37}) in Section \ref{moon} we obtain:
\begin{multline*} 
	p'-S'= de^{-i\Omega_5 t} +
		\frac
		{b_1 
			\{
				1 - 
					(
						1 - \cos \chi_0
					) \Omega_5/ \Omega_1
			\} }
		{\Omega_5/\Omega_1 - 1}
		e^{-i\Omega_1 t} +\\
		\frac
		{b_2 
			\{
				1 - 
					(
						1 - \cos \chi_0
					)
					 \Omega_5/\Omega_2
			\} }
		{\Omega_5/ \Omega_2 - 1}
		e^{-i\Omega_2 t} +
		\frac
		{b_3 
			\{
				1 - 
					(
						1 - \cos \chi_0
					) 
					\Omega_5/\Omega_3
			\} }
		{ \Omega_5/\Omega_3 - 1}
		e^{-i\Omega_3 t} 
\end{multline*} 
 The numerators of the fractions can be evaluated numerically using the figures in Section \ref{spin} giving:
 \begin{equation*}
	p'-S'= de^{-i\Omega_5 t} +
		\frac
		{0.772 b_1   }
		 {\Omega_5/ \Omega_1 - 1}
		e^{-i\Omega_1 t} +
		\frac
		{0.304 b_2  }
		{ \Omega_5/ \Omega_2 - 1}
		e^{-i\Omega_2 t} +
		\frac
		{ 0.842 b_3  }
		{ \Omega_5/ \Omega_3 - 1}
		e^{-i\Omega_3 t} 
\end{equation*} 
The factors 0.772, 0.304 and 0.872 are absent from (\ref{eq:2.37}): this is evidently the effect of the tilt of the Earth’s spin axis not being small.

\end{document}